\documentclass[11pt,graphicx,subfigure,axodraw,date]{article}
\usepackage{amsfonts}
\usepackage{amssymb}
\usepackage[usenames,dvipsnames]{color}
\usepackage{graphicx}
\usepackage{amsmath}
\usepackage{amsfonts}
\usepackage{amssymb}
\usepackage{cleveref}
\usepackage{float}
\restylefloat{table}
\usepackage{chngpage}
\def\rmuu{\gamma^{\mu}}
\def\rmud{\gamma_{\mu}}
\def\PL{{1-\gamma_5\over 2}}
\def\PR{{1+\gamma_5\over 2}}
\def\sinW2{\sin^2\theta_W}
\def\AEM{\alpha_{EM}}
\def\mul{M_{\tilde{u} L}^2}
\def\mur{M_{\tilde{u} R}^2}
\def\mdl{M_{\tilde{d} L}^2}
\def\mdr{M_{\tilde{d} R}^2}
\def\mz2{M_{z}^2}
\def\c2b{\cos 2\beta}
\def\au{A_u}
\def\ad{A_d}
\def\cob{\cot \beta}
\def\v#1{v_#1}
\def\tb{\tan\beta}
\def\epem{$e^+e^-$}
\def\KK{$K^0$-$\overline{K^0}$}
\def\wi{\omega_i}
\def\xj{\chi_j}
\def\Wmu{W_\mu}
\def\Wnu{W_\nu}
\def\m#1{{\tilde m}_#1}
\def\mH{m_H}
\def\mw#1{{\tilde m}_{\omega #1}}
\def\mx#1{{\tilde m}_{\chi^{0}_#1}}
\def\mc#1{{\tilde m}_{\chi^{+}_#1}}
\def\mwi{{\tilde m}_{\omega i}}
\def\mxi{{\tilde m}_{\chi^{0}_i}}
\def\mci{{\tilde m}_{\chi^{+}_i}}

\def\ch{{\tilde\chi^{+}_1}}
\def\c2{{\tilde\chi^{+}_2}}

\def\tt{{\tilde\theta}}

\def\tp{{\tilde\phi}}

\def\mz{M_z}
\def\sw{\sin\theta_W}
\def\cw{\cos\theta_W}
\def\cb{\cos\beta}
\def\rwi{r_{\omega i}}
\def\rxj{r_{\chi j}}
\def\rfp{r_f'}
\def\Kik{K_{ik}}
\def\Fq2{F_{2}(q^2)}
\def\f{\({\cal F}\)}
\def\d1{{\f(\tilde c;\tilde s;\tilde W)+ \f(\tilde c;\tilde \mu;\tilde W)}}
\def\tw{\tan\theta_W}
\def\sec2w{sec^2\theta_W}
\begin{document}
\baselineskip 18pt
\def\today{\ifcase\month\or
 January\or February\or March\or April\or May\or June\or
 July\or August\or September\or October\or November\or December\fi
 \space\number\day, \number\year}
\def\thebibliography#1{\section*{References\markboth
 {References}{References}}\list
 {[\arabic{enumi}]}{\settowidth\labelwidth{[#1]}
 \leftmargin\labelwidth
 \advance\leftmargin\labelsep
 \usecounter{enumi}}
 \def\newblock{\hskip .11em plus .33em minus .07em}
 \sloppy
 \sfcode`\.=1000\relax}
\let\endthebibliography=\endlist
\def\lsim{\ ^<\llap{$_\sim$}\ }
\def\gsim{\ ^>\llap{$_\sim$}\ }
\def\r2{\sqrt 2}
\def\beq{\begin{equation}}
\def\eeq{\end{equation}}
\def\beqn{\begin{eqnarray}}
\def\eeqn{\end{eqnarray}}
\def\rmuu{\gamma^{\mu}}
\def\rmud{\gamma_{\mu}}
\def\PL{{1-\gamma_5\over 2}}
\def\PR{{1+\gamma_5\over 2}}
\def\sinW2{\sin^2\theta_W}
\def\AEM{\alpha_{EM}}
\def\mul{M_{\tilde{u} L}^2}
\def\mur{M_{\tilde{u} R}^2}
\def\mdl{M_{\tilde{d} L}^2}
\def\mdr{M_{\tilde{d} R}^2}
\def\mz2{M_{z}^2}
\def\c2b{\cos 2\beta}
\def\au{A_u}
\def\ad{A_d}
\def\cob{\cot \beta}
\def\v#1{v_#1}
\def\tb{\tan\beta}
\def\epem{$e^+e^-$}
\def\KK{$K^0$-$\bar{K^0}$}
\def\wi{\omega_i}
\def\xj{\chi_j}
\def\Wmu{W_\mu}
\def\Wnu{W_\nu}
\def\m#1{{\tilde m}_#1}
\def\mH{m_H}
\def\mw#1{{\tilde m}_{\omega #1}}
\def\mx#1{{\tilde m}_{\chi^{0}_#1}}
\def\mc#1{{\tilde m}_{\chi^{+}_#1}}
\def\mwi{{\tilde m}_{\omega i}}
\def\mxi{{\tilde m}_{\chi^{0}_i}}
\def\mci{{\tilde m}_{\chi^{+}_i}}
\def\mz{M_z}
\def\sw{\sin\theta_W}
\def\cw{\cos\theta_W}
\def\cb{\cos\beta}
\def\rwi{r_{\omega i}}
\def\rxj{r_{\chi j}}
\def\rfp{r_f'}
\def\Kik{K_{ik}}
\def\Fq2{F_{2}(q^2)}
\def\f{\({\cal F}\)}
\def\d1{{\f(\tilde c;\tilde s;\tilde W)+ \f(\tilde c;\tilde \mu;\tilde W)}}
\def\tw{\tan\theta_W}
\def\sec2w{sec^2\theta_W}
\def\ch{{\tilde\chi^{+}_1}}
\def\c2{{\tilde\chi^{+}_2}}

\def\tt{{\tilde\theta}}

\def\tp{{\tilde\phi}}

\def\mz{M_z}
\def\sw{\sin\theta_W}
\def\cw{\cos\theta_W}
\def\cb{\cos\beta}
\def\rwi{r_{\omega i}}
\def\rxj{r_{\chi j}}
\def\rfp{r_f'}
\def\Kik{K_{ik}}
\def\Fq2{F_{2}(q^2)}
\def\f{\({\cal F}\)}
\def\d1{{\f(\tilde c;\tilde s;\tilde W)+ \f(\tilde c;\tilde \mu;\tilde W)}}

\def\b{${\cal{B}}(\mu\to {e} \gamma)$~}

\def\tw{\tan\theta_W}
\def\sec2w{sec^2\theta_W}

\newcommand{\ti}[1]{{\color{blue}{#1}}}
\newcommand{\pg}[1]{{\color{green}{#1}}}
\newcommand{\pr}[1]{{\colorx{red}{#1}}}
\def\flv{flavor~violating~}

\def\fvq{flavor~violating~quark~decays~of~the~Higgs~boson~}

\def\non{\nonumber\\}

\def\sb{  b \bar s +\bar b s }
\def\db{  b \bar d +\bar b d }

\begin{titlepage}

\begin{center}
{\large {\bf
 Flavor violating top decays and flavor violating quark decays of the Higgs boson
 }}
 
\renewcommand{\thefootnote}
{\fnsymbol{footnote}}
Tarek Ibrahim$^{a}$\footnote{Email: tibrahim@zewailcity.edu.eg},
Ahmad Itani$^{b}$\footnote{Email: ahmad.it@gmail.com},
   Pran Nath$^{c}$\footnote{Email: nath@neu.edu}, Anas Zorik$^{a}$\footnote{Email: p-azorik@zewailcity.edu.eg}
\end{center}

\date{Feb 14, 2015}

\noindent
{$^{a}$University of Science and Technology, Zewail City of Science and Technology,}\\
{ 6th of October City, Giza 12588, Egypt\footnote{Permanent address:  Department of  Physics, Faculty of Science,
University of Alexandria, Alexandria, Egypt}\\
} 
{$^{b}$Department of Physics, Beirut Arab University, Beirut 11-5020,Lebanon} \\
{$^{c}$Department of Physics, Northeastern University,
Boston, MA 02115-5000, USA} \\

\centerline{\bf Abstract}

In the standard model \flv decays of the top quark and of  the Higgs boson are highly suppressed. 
Further, the \flv decays of the top and of the Higgs are also small in MSSM and not observable in 
current or in near future experiment. 
 In this work we show that much larger branching ratios for these decays can be achieved in 
 an extended MSSM model with an additional vector like quark generation. 
 Specifically we show that in the extended model one can achieve 
 branching ratios for  $t\to h^0 c$ and $t\to h^0 u$ as large as the current experimental upper limits  
given by the ATLAS and the CMS Collaborations.  
We also analyze the flavor violating quark decay of the Higgs boson, i.e.,  $h^0\to \sb$ and 
$h^0\to b\bar d + \bar b d$. Here again one finds that the branching ratio for these
decays can be as large as $O(1)\%$. The analysis is done with inclusion of the CP phases in the Higgs sector, and 
the effect of CP phases on the branching ratios is investigated.  Specifically
the Higgs sector spectrum and mixings are computed involving  quarks and mirror quarks, squarks  and 
mirror squarks in the loops consistent with the Higgs boson mass constraint.
 The resulting effective Lagrangian with inclusion of the vector like quark generation 
 induce flavor violating decays at the tree level.  The test of the branching ratios predicted could
come with  further data from LHC13 and such branching ratios could also be accessible at future colliders
such as the Higgs factories where the Higgs couplings to fermions will be determined with greater
precision. 

\medskip
\end{titlepage}
\medskip

\section{Introduction \label{sec1}}
The flavor violating  decays provide a window to new physics beyond the standard model. 
Thus in the standard model the \flv top decay $t\to h^0 c$ has a branching ratio which is rather small, i.e., 
$O(10^{-15})$ \cite{ehs,mps,jaa,zm}. In the two Higgs doublet model the branching ratio is predicted to be in the range  $10^{-5} -10^{-3}$
\cite{14,15,16,17,18,19,20}.
Currently  experimental results  on $t\to  h^0 + c$  and on
 $t\to h^0+ u$  from the ATLAS collaboration~\cite{Aad:2015pja} and from the CMS 
 Collaboration ~\cite{Khachatryan:2016atv} are 
 as follows (for a review of experiment see ~\cite{Araque:2016odm}):
 From the ATLAS Collaboration ~\cite{Aad:2015pja} one has for  the  branching ratios  
\begin{align}
BR(t\to h^0+c) & < 0.56 ~(95\%{~\rm CL})\,, \nonumber\\
BR(t\to h^0+u)  & < 0.61 ~(95\% {~\rm CL})\,,
\label{1}
\end{align}
and from the  CMS Collaboration~\cite{Khachatryan:2016atv} one has 
\begin{align}
BR(t\to h^0+c)  & < 0.40 ~(95\% {~\rm CL})\,, \nonumber\\
BR(t\to h^0+u)  & < 0.55 ~(95\% {~\rm CL})\,.
\label{2}
\end{align}

The \flv decays of the Higgs boson offer another window to the discovery of new physics. 
In a previous work we analyzed the \flv Higgs decays into leptons using an extension of MSSM
with a vector like leptonic generation~\cite{Fathy:2016vli}.
Experimental limits on such decays exist from 
the ATLAS\cite{Aad:2015gha}  and from the CMS~\cite{Khachatryan:2015kon} 
Collaborations.  In~\cite{Fathy:2016vli}.
it  was shown that in the extended MSSM model one could achieve
up to  $O(1)\%$ 
branching ratio for the \flv leptonic decay $h^0\to \tau\mu$.  
Here we carry out a similar analysis with inclusion of a vector like quark generation
to analyze the flavor violating Higgs couplings to quarks 
which we utilize to compute the \flv decays of the top $t\to h^0 c$ and $t\to h^0 u$. 
Other significant channels for the observation of \flv process are the decays 
$h^0 \to \sb$ and  $h^0\to \db$.
In the standard model the branching ratio for 
 such a processes is order  $10^{-7}$~\cite{Benitez-Guzman:2015ana}  or less
and unobservable. 
In the framework of SUGRA/MSSM model the branching ratio is still small, i.e., 
 $O(10^{-4})$~\cite{Gomez:2015duj,Barenboim:2015fya,Arhrib:2006vy}. 
 In the extended MSSM model discussed here, we find that the branching ratios $BR(h^0\to \sb)$ 
 and  $BR(h^0\to \db)$ 
 can be 
 as large as $O(1)\%$  which is several orders of magnitude larger than in  MSSM and possibly within 
 reach of current and future experiment.
 More data is expected in the near future which makes an investigation of the \flv decays  of the top and of the 
Higgs a timely topic of investigation. \\

The outline of the rest of the paper is as follows: In section \ref{sec2} a discussion of the extended MSSM model
with a vector like generation is given. Here it is shown that the quark couplings involve flavor violating vertices.
In section 3 an analysis of the \flv top decays $t\to h^0 c$ and $t\to h^0 u$ is given. 
In section \ref{sec4}  the \flv Higgs decays $h^0\to  \sb$ and $h^0\to \db$
are discussed.
In sec \ref{sec5} a numerical analysis of sizes of the branching ratios of the \flv processes is given.
Conclusions are given in sec \ref{sec6}. Further details of the squark mass squared matrices including the vectorlike
squarks are given in the appendix A and a brief discussion of CP even -CP odd Higgs mixing is given in appendix B.

\section{The Model and Notation\label{sec2}}
Here we  describe the model briefly and further details are given in the appendix A and appendix B.  
The model we consider is an
extension of MSSM with an additional vectorlike multiplet. 
Vectorlike multiplets appear in a variety of settings which include grand unified models,
string and D brane models~\cite{vectorlike,Babu:2008ge,Liu:2009cc,Martin:2009bg}. 
Further, they are anomaly free.
Several analyses have recently
appeared which utilize such multiplets
\cite{Ibrahim:2008gg,Ibrahim:2010va,Ibrahim:2010hv,Ibrahim:2011im,Ibrahim:2012ds,Aboubrahim:2013gfa,Aboubrahim:2015zpa,Ibrahim:2015hva,Ibrahim:2014oia,Aboubrahim:2016xuz}.
 Here we focus on the quark sector where the vectorlike multiplet consists of a
 fourth generation of quarks and their mirror quarks.
 Thus the quark sector of the extended MSSM model has the matter content given by 

\begin{align}
q_{iL}\equiv
 \left(\begin{matrix} t_{i L}\cr
 ~{b}_{iL}  \end{matrix} \right)  \sim \left(3,2,\frac{1}{6}\right) \ ;  ~~ ~t^c_{iL}\sim \left(3^*,1,-\frac{2}{3}\right)\ ;
 ~~~ b^c_{i L}\sim \left(3^*,1,\frac{1}{3}\right)\ ;
  ~~~i=1,2,3,4.
\label{2a}
\end{align}

\begin{align}
Q^c\equiv
 \left(\begin{matrix} B_{ L}^c \cr
 T_L^c\end{matrix}\right)  \sim \left(3^*,2,-\frac{1}{6}\right)\ ;
~~  T_L \sim  \left(3,1,\frac{2}{3}\right)\ ;  ~~   B_L \sim \left(3^*,1,-\frac{1}{3}\right).
\label{3a}
\end{align}
The numbers in the braces in Eq.(3) and Eq.(4)
show  the properties  under $SU(3)_C\times SU(2)_L\times U(1)_Y$
where the first two entries label the representations for $SU(3)_C$ and $SU(2)_L$ and the last one
gives the value of the hypercharge normalized so  that $Q=T_3+Y$.
We allow the mixing of the vectorlike generation with the first three generations. 
We display now some relevant features of the model.  In the up quark sector we
choose a basis as follows

\begin{gather}
\bar\xi_R^T= \left(\begin{matrix}\bar t_{ R} & \bar T_R & \bar c_{ R}
&\bar u_{R} &\bar t_{4R} \end{matrix}\right),~~
\xi_L^T= \left(\begin{matrix} t_{ L} &  T_L &  c_{ L}
& u_{ L} &\bar t_{4L}\end{matrix}\right)\,.
\label{basis-xi}
\end{gather}
and we write the mass term  so that

\beq
-{\cal L}^u_m= \bar\xi_R^T (M_u) \xi_L
+\text{h.c.}\,.
\eeq
The superpotential  of the model (as shown in appendix A)
  leads to the up-quark  mass matrix $M_u$  where

\beqn
M_u=
 \left(\begin{matrix} y'_1 v_2/\sqrt{2} & h_5 & 0 & 0&0 \cr
 -h_3 & y_2 v_1/\sqrt{2} & -h_3' & -h_3''&-h_6 \cr
0&h_5'&y_3' v_2/\sqrt{2} & 0 &0\cr
0 & h_5'' & 0 & y_4' v_2/\sqrt{2}&0 \cr
0&h_8&0&0&y_5'v_2/\sqrt{2}\end{matrix}\right)\,.
\label{Mup}
\eeqn
This mass matrix is not hermitian and a  bi-unitary transformation is needed  to diagonalize it.
Thus one has
\beq
D^{u \dagger}_R (M_u) D^u_L=\text{diag}(m_{u_1},m_{u_2},m_{u_3}, m_{u_4},  m_{u_5} )\,.
\label{7a}
\eeq
Under the bi-unitary transformations the basis vectors transform so that
\beqn
 \left(\begin{matrix} t_{R}\cr
 T_{ R} \cr
c_{R} \cr
u_{R} \cr
t_{4R}
\end{matrix}\right)=D^{u}_R \left(\begin{matrix} u_{1_R}\cr
 u_{2_R}  \cr
u_{3_R} \cr
u_{4_R}\cr
u_{5_R}
\end{matrix}\right), \  \
\left(\begin{matrix} t_{L}\cr
 T_{ L} \cr
c_{L} \cr
u_{L}\cr
t_{4L}
\end{matrix} \right)=D^{u}_L \left(\begin{matrix} u_{1_L}\cr
 u_{2_L} \cr
u_{3_L} \cr
u_{4_L}\cr
u_{5_L}
\end{matrix}\right) \ .
\label{8}
\eeqn

A similar analysis can be carried out for the down quarks. Here we choose the basis set as
\begin{gather}
\bar\eta_R^T= \left(\begin{matrix}\bar{b}_R & \bar B_R & \bar{s}_R
&\bar{d}_R
&\bar{b}_{4R}
\end{matrix}\right),
~~\eta_L^T= \left(\begin{matrix} {b_ L} &  B_L &  {s_ L}
& {d_ L}
&{b_{4L}}
 \end{matrix}\right)\,.
\label{basis-eta}
\end{gather}

In this basis  the down quark mass terms are given by
\beq
-{\cal L}^d_m=
\bar\eta_R^T(M_{d}) \eta_L
+\text{h.c.},
\eeq
where using the interactions of  appendix A,  $M_d$ has the following form
\beqn
M_d=\left(\begin{matrix} y_1 v_1/\sqrt{2} & h_4 & 0 & 0  & 0\cr
 h_3 & y'_2 v_2/\sqrt{2} & h_3' & h_3'' &h_6\cr
0&h_4'&y_3 v_1/\sqrt{2} & 0&0 \cr
0 & h_4'' & 0 & y_4 v_1/\sqrt{2}&0\cr
0& h_7 & 0 &0 &y_5 v_1/\sqrt{2}
\end{matrix} \right)\ .
\label{7bb}
\label{Mdown}
\eeqn
In general   the parameters $h_3, h_4, h_5, h_3', h_4',h_5',  h_3'', h_4'',h_5'', h_6, h_7, h_8$ 
appearing in Eqs. (\ref{Mup}) and (\ref{Mdown})
can be complex and we define their phases
so that

\beqn
h_k= |h_k| e^{i\chi_k}, ~~h_k'= |h_k'| e^{i\chi_k'}, ~~~h_k''= |h_k''| e^{i\chi_k''}\,.
\label{mix}
\eeqn
The squark sector of the model contains a variety of terms including F -type, D-type,  as well as
soft mixings terms involving squarks and mirror squarks. The details of these contributions to squark mass
square matrices are  discussed in  appendix A. In addition to the CP phases arising from the {mixing parameters} 
as given by Eq. (\ref{mix}) there are CP phases arising from the soft parameters 
as discussed in appendix A.  In general these phases can be large. The CP phases in general contribute to the
EDM of the quarks and the leptons. Compatiblity with experiment can be achieved in a variety of ways 
such as via mass suppression~\cite{Nath:1991dn,Kizukuri:1992nj}  
 or the cancellation mechanism~\cite{Ibrahim:1998je}~\cite{Ibrahim:1997gj,Falk:1998pu,Brhlik:1998zn,Ibrahim:1999af}
 (for a review see~\cite{Ibrahim:2007fb}).

We note that the up quark matrix given by Eq. (\ref{Mup}) and the down quark matrix given by 
Eq. (\ref{Mdown}) contain  off diagonal elements in the flavor space. Additionally Eq. (23) and  Eq. (28)
contain flavor mixings.  It is the presence of these mixings that lead to flavor violating decays of the
top quark and  the flavor violating decays of the Higgs boson.
This was done with inclusion of CP violating phases in the Higgs sector. An analysis of the effects of CP phases
on the Higgs boson masses and mixings with inclusion of the vector like generation was given in 
~\cite{Ibrahim:2016rcb} and we utilize the work of that analysis here. 
(For previous work on the effects of CP phases
on Higgs boson massess and mixings see  
\cite{Pilaftsis:1998pe,Pilaftsis:1998dd,Pilaftsis:1999qt,Demir:1999hj,Choi:2000wz,Carena:2000yi,Ibrahim:2000qj,Ibrahim:2002zk,FeynHiggs,Lee:2012wa}).
The inclusion of CP phases affects the Higgs boson masses
and mixings. Specifically  the mass eigenstates of the neutral Higgs bosons are no longer CP eigenstates 
and further their couplings to quarks and leptons are dependent on the phases. In this work we further
show that the coupling of the neutral Higgs bosons with the quarks allow flavor violating decays
  even at the tree level. Specifically we have analyzed  the loop corrections to the scalar potential in the Higgs sector using the super trace technique. We have produced the corrected Higgs mass$^2$ matrix and diagonalized it  
\beq
Y M^2 Y^T = diag(M^2_{H_1}, M^2_{H_2}, M^2_{H_3} )\,.
\eeq

A brief discussion of the CP phases on  the CP even-CP odd Higgs mixing is given in  appendix B.

\section{Flavor violating top decays: $t\to h^0 c$ and $t\to h^0 u$\label{sec3}}
In this section we compute the flavor violating decays of the top quark, i.e., 

\begin{align}
t\to h^0 c,  ~~t\to h^0 u\,.
\label{topdecays}
\end{align}
As mentioned in section \ref{sec1} experimental upper limits from the ATLAS ~\cite{Aad:2015pja} 
and from  and the CMS~\cite{Khachatryan:2015kon}  Collaborations 
exist on these decays 
as exhibited in Eq.(1)  and Eq.(2). In the analysis presented here we will show that branching ratios for the processes
Eq. (15) close to the upper limit exhibited in Eq. (1) can be achieved. 

Using the superpotential Eq.(23), one can write the interaction between the mass eigen states of the Higgs bosons $H_k$ and the quark mass eigen states so that 
\beqn
{\cal{L}}= \bar{d}_i (\epsilon_{ijk}+\gamma_5 \epsilon'_{ijk})d_j H_k
+\bar{u}_i (\eta_{ijk}+\gamma_5 \eta'_{ijk})u_j H_k\,,
\label{interaction}
\eeqn
where the couplings are given by
\beqn
\epsilon_{ijk}=-\frac{1}{2\sqrt{2}}\{\phi_{ij}(Y_{k2}+iY_{k3}\cos\beta)+\phi^*_{ji}(Y_{k2}-iY_{k3}\cos\beta)\nonumber\\
+\alpha_{ij}(Y_{k1}+iY_{k3}\sin\beta)+\alpha^*_{ji}(Y_{k1}-iY_{k3}\sin\beta)\},\nonumber\\
\epsilon'_{ijk}=-\frac{1}{2\sqrt{2}}\{-\phi_{ij}(Y_{k2}+iY_{k3}\cos\beta)+\phi^*_{ji}(Y_{k2}-iY_{k3}\cos\beta)\nonumber\\
-\alpha_{ij}(Y_{k1}+iY_{k3}\sin\beta)+\alpha^*_{ji}(Y_{k1}-iY_{k3}\sin\beta)\},\nonumber\\
\eta_{ijk}=-\frac{1}{2\sqrt{2}}\{\phi'_{ij}(Y_{k2}+iY_{k3}\cos\beta)+\phi^{'*}_{ji}(Y_{k2}-iY_{k3}\cos\beta)\nonumber\\
+\alpha'_{ij}(Y_{k1}+iY_{k3}\sin\beta)+\alpha^{'*}_{ji}(Y_{k1}-iY_{k3}\sin\beta)\},\nonumber\\
\eta'_{ijk}=-\frac{1}{2\sqrt{2}}\{-\phi'_{ij}(Y_{k2}+iY_{k3}\cos\beta)+\phi^{'*}_{ji}(Y_{k2}-iY_{k3}\cos\beta)\nonumber\\
-\alpha'_{ij}(Y_{k1}+iY_{k3}\sin\beta)+\alpha^{'*}_{ji}(Y_{k1}-iY_{k3}\sin\beta)\}\,,
\eeqn
and the parameters $\phi_{ij}$, $\alpha_{ij}$, $\phi'_{ij}$ and $\alpha'{ij}$ are given by
\begin{align}
\phi_{ij}& =y'_2 D^{d*}_{R_{2i}}  D^{d}_{L_{2j}}\,, \nonumber\\
\alpha_{ij}&=y_1 D^{d*}_{R_{1i}}  D^{d}_{L_{1j}}
+y_3 D^{d*}_{R_{3i}}  D^{d}_{L_{3j}}+
y_4 D^{d*}_{R_{4i}}  D^{d}_{L_{4j}}+
y_5 D^{d*}_{R_{5i}}  D^{d}_{L_{5j}}\,,
\nonumber\\
\phi'_{ij}&=y'_1 D^{u*}_{R_{1i}}  D^{u}_{L_{1j}}
+y'_3 D^{u*}_{R_{3i}}  D^{u}_{L_{3j}}+
y'_4 D^{u*}_{R_{4i}}  D^{u}_{L_{4j}}+
y'_5 D^{u*}_{R_{5i}}  D^{u}_{L_{5j}}\,,
\nonumber\\
\alpha'_{ij}&=y_2 D^{u*}_{R_{2i}}  D^{u}_{L_{2j}}\,.
\end{align}

Using the interaction (\ref{interaction}), we calculate the decay widths of the top quark into the lightest Higgs and the flavors $c$ and $u$ to be
\begin{align}
\Gamma(t\rightarrow H_1 +c)&= \frac{1}{16\pi m^3_t}\sqrt{(m^2_t+m^2_c -M^2_{H_1})^2-4m^2_t m^2_c}\nonumber\\
&\{(m_t+m_c)^2 |\eta_{311}|^2
+(m_t -m_c)^2 |\eta'_{311}|^2 -M^2_{H_1} (|\eta_{311}|^2+|\eta'_{311}|^2)
\}\,,\nonumber\\
\Gamma(t\rightarrow H_1 +u)&= \frac{1}{16\pi m^3_t}\sqrt{(m^2_t+m^2_u -M^2_{H_1})^2-4m^2_t m^2_u}\nonumber\\
&\{(m_t+m_u)^2 |\eta_{411}|^2
+(m_t -m_u)^2 |\eta'_{411}|^2 -M^2_{H_1} (|\eta_{411}|^2+|\eta'_{411}|^2)
\}\,.
\end{align}
To calculate the branching ratio of the top quark to the above channels, we just need to divide the partial widths for these 
top decays  by the total width of the top quark which is $1.41^{+.19}_{-0.15}$ GeV. In the analysis here we take 
the center value of $1.41$ GeV for the width. 

\section{Flavor violating Higgs boson decays: $h^0\to  \sb$ and $h^0\to \db$
\label{sec4}}
We proceed now to discuss the \flv quark decays of the Higgs, and specifically
$h^0\to \sb$.
In  future data from collider experiments is likely to either discover such decays or put more stringent 
constraints 
on them. This may happen with LHC13 data after the high luminosity upgrade. Further, \flv decays would be explored
at Higgs factories which are under active consideration such as the International Linear Collider
(ILC- Japan), Circular Electron-Positron Collider (CEPC-China) or Future Circular Collider-ee (FCC-ee: CERN).
Thus \flv decays of the Higgs are  of considerable interest.

Using
 the interaction (\ref{interaction}), we calculate the decay widths of the lightest Higgs boson  into $b+\bar{s}$ and $\bar{b} + s$ to be

\begin{align}
\Gamma(H_1\rightarrow b+ \bar{s})&= \frac{3}{8\pi M^3_{H_1}}\sqrt{(m^2_b+m^2_s -M^2_{H_1})^2-4m^2_b m^2_s}\nonumber\\
&\{(|\epsilon_{311}|^2+|\epsilon'_{311}|^2)(M^2_{H_1}-m^2_b-m^2_s)-(|\epsilon_{311}|^2-|\epsilon'_{311}|^2)
(2 m_b m_s)
\}\,,\nonumber\\
\Gamma(H_1\rightarrow \bar{b}+s)&= \frac{3}{8\pi M^3_{H_1}}\sqrt{(m^2_b+m^2_s -M^2_{H_1})^2-4m^2_b m^2_s}\nonumber\\
&\{(|\epsilon_{131}|^2+|\epsilon'_{131}|^2)(M^2_{H_1}-m^2_b-m^2_s)-(|\epsilon_{131}|^2-|\epsilon'_{131}|^2)
(2 m_b m_s)
\}\,.
 \end{align}
To calculate the branching ratio of the lightest Higgs to the above channels, we just need to divide the
partial decay widths  by the total width of the Higgs boson.
Thus the  flavor  violating branching ratio of $H_1$ into $ \sb$  is given by 

\beqn
BR(H_1 \rightarrow  \sb )= \frac{\Gamma (H_1 \rightarrow b \bar s)    + \Gamma (H_1 \rightarrow \bar b s) }{\Gamma (H_1 \rightarrow \bar{b} s)+\Gamma (H_1 \rightarrow \bar{s} b)+\sum_{i} \Gamma (H_1 \rightarrow \bar{f_i} f_i)
+ \Gamma_{H_1DB}}\,,
\label{eq45}
\eeqn
where $f_i$ stand for fermionic particles  that have coupling with the Higgs boson and have a mass  less than half the Higgs boson mass and $\Gamma_{H_1DB}$  is the decay width into diboson states which 
include $gg,  \gamma\gamma, \gamma Z, ZZ, WW$. 
Thus the computation of  the branching ratios of Eq. (\ref{eq45}) involve the decay widths
\beqn
\Gamma_i (H_i \rightarrow \bar{f} f)_{f=b,d,s}=\frac{3g^2 m^2_f }{32\pi m^2_{W} \cos^2\beta} M_i \{ |Y_{i1}|^2(1-\frac{4m^2_f}{M^2_i})^{3/2} + |Y_{i3}|^2 \sin^2\beta (1-\frac{4m^2_f}{M^2_i})^{1/2}
\}\,,\nonumber\\
\Gamma_i (H_i \rightarrow \bar{f} f)_{f=\tau, \mu, e}=\frac{g^2 m^2_f }{32\pi m^2_{W} \cos^2\beta} M_i \{ |Y_{i1}|^2(1-\frac{4m^2_f}{M^2_i})^{3/2} + |Y_{i3}|^2 \sin^2\beta (1-\frac{4m^2_f}{M^2_i})^{1/2}
\}\,,\nonumber\\
\Gamma_i (H_i \rightarrow \bar{f} f)_{f=u,c}=\frac{3g^2 m^2_f }{32\pi m^2_{W} \sin^2\beta} M_i \{ |Y_{i2}|^2(1-\frac{4m^2_f}{M^2_i})^{3/2} + |Y_{i3}|^2 \cos^2\beta (1-\frac{4m^2_f}{M^2_i})^{1/2}
\}\,.
\label{eq46}
\eeqn
An identical analysis for $H_1\to \db$ holds with $s$ replaced with $d$.
In the decays of the lightest Higgs 
into $ZZ$ and $WW$, these states are off shell and the on-shell final states are  dominantly 
four fermions arising from the decay of the $Z$ and $W$ bosons. We note that $\sb$ and $\db$ 
final states do not originate from any of the $ZZ$ and $WW$  
 diboson decay modes.
Further,  the Higgs boson observed at $\sim 125$ GeV ~\cite{Chatrchyan:2012ufa,Aad:2012tfa}
 is effectively in the decoupling limit. Thus we approximate
the diboson decay widths as given by the standard model.

\section{Discussion of numerical results\label{sec5}}

We discuss now the numerical analysis of the flavor violating decays of the top: $t\to  h^0 + c$  and 
 $t\to h^0 + u$, and the flavor violating Higgs decay $h^0\to \sb$ { and $h^0\to \db$}. 
 As a first step we diagonalize the $5\times 5$ up  and down quark and mirror quark mass matrices.
 In the diagonalzation the parameters are chosen so as produce the masses of the three generation of 
 up and down quarks as given by the PDG~\cite{Olive:2016xmw}. Further, the inputs are chosen to generate the masses of the 
 remaining  quarks  and mirror quarks to be consistent with the lower bounds given by PDG~\cite{Olive:2016xmw} for heavy quarks.  
 As discussed in section \ref{sec2} we include loop corrections to the Higgs boson potential which
generate mixings of the CP even-CP odd Higgs leading to the mass eigenstates $H_1 H_2 H_3$
 which are not eigenstates of CP. The analysis of the loop corrections to the Higgs involves masses of the 
 squarks and the mirror squarks which are given in appendix {B}.  In the numerical analysis of the loop corrections
we make the following simplifying assumption:
  $m^{u^2}_0=M^2_{\tilde T}=M^2_{\tilde t_1}=M^2_{\tilde t_2}=M^2_{\tilde t_3}=M^2_{\tilde t_4}$ and $m^{d^2}_0=M^2_{\tilde 1 L}=M^2_{\tilde B}=M^2_{\tilde b_1}=M^2_{\tilde Q}=M^2_{\tilde 2 L}=M^2_{\tilde b_2}=M^2_{\tilde 3 L}=M^2_{\tilde b_3}=M^2_{\tilde 4 L}=M^2_{\tilde b_4}$. {{and}} $m^u_0=m^d_0=m_0$.
 Additionally the trilinear couplings are chosen so that: $A^u_0=A_t=A_T=A_c=A_u=A_{4t}$ and $A^d_0=A_b=A_B=A_s=A_d=A_{4b}$.

 In table \ref{table:1}  (see caption) we exhibit  the inputs for the matrices of Eqs.~\ref{Mup}  and ~\ref{Mdown} which
 lead to the masses of the three generations quarks as given by the PDG~\cite{Olive:2016xmw} and produce masses
 for the extra quarks and mirror quarks. The masses of these extra quarks and mirror quarks are exhibited
 in table \ref{table:1} and they are consistent with the lower limits of the heavy quarks given by PDG~\cite{Olive:2016xmw}.
 We utilize the inputs of table \ref{table:1}  in the computation of the flavor violating branching ratios of the top 
 and the Higgs given in table \ref{table:2}. Here, however, we need to specify also the soft parameters.
 These are chosen so they provide the desired loop correction to the Higgs boson mass  to be 
   consistent with the experimental value of $\sim 125$ GeV.  The inputs are given in the caption of 
   table \ref{table:2}.  For the flavor violating top decays $t\to h^0 c$ and $t\to h^0 u$, the branching ratios
   are orders of magnitude larger than achievable in the standard model or in the MSSM, and    
   come close to the upper limits given by ATLAS~\cite{Aad:2015pja}(see Eq.(1)) 
   and CMS~\cite{Khachatryan:2016atv}(see Eq.(2)).  A similar result holds for the flavor violating quark 
   decay of the Higgs into $\sb$ { and $\db$}.    Again in the standard model and in MSSM, the branching ratio for
   the flavor violating decay of the Higgs is orders of magnitude smaller than a precent.  However, 
   in the extended MSSM model discussed here, the branching ratio can be as large as  $O(1)\%$. A branching 
   ratio of this size could be tested with more data from  LHC13 and possibly at future 
   colliders such as the Higgs factories.
   
\begin{table}[H]
\begin{center}
\begin{tabular}{l l  l  l}
 \hline\hline
 \multicolumn{2}{c}{ Heavy Up Quarks}       & \multicolumn{2}{c}{ Heavy Down Quarks} \\
 \hline
 Mirror Up Quark           & $m_{t'}$ = $803$          & Mirror Down Quark           & $m_{b'}$ = $817$          \\
 Fourth Generation Up Quark & $m^{\rm up}_{4}$ = $917$ & Fourth Generation Down Quark & $m^{\rm down}_{4}$ = $1044$ \\
 \hline\hline
 \end{tabular}
 \caption{An exhibition of masses of the vectorlike quarks  from  diagonalization of the matrices of Eqs.(\ref{Mup})
 and (\ref{Mdown}) consistent with the current lower bounds on exotic quarks.
 The parameters used are $|h_3|=1.6$, $|h'_3|=6.34\times10^{-2}$, $|h''_3|=1.97\times10^{-2}$, $|h_4|=485$, $|h'_4|=500$,
 $|h''_4|=410$, $|h_5|=22$, $|h'_5|=545$, $|h''_5|=130$, $|h_6|=550$, $|h_7|=280$, $|h_8|=450$, $\chi_3=0.9$,
 $\chi'_3=1\times10^{-3}$, $\chi''_3=4\times10^{-3}$, $\chi_4=\chi'_4=2.1$, $\chi''_4=0.6$, $\chi_5=0.99$,
 $\chi'_5=2.7$, $\chi''_5=2.4$, $\chi_6=0.01$, $\chi_7=3.1$, $\chi_8=0.01$. The input of the diagonal elements are 
 ($173.21$, $420$, $1.7$, $2.3\times10^{-3}$, $700$) for Eq.(\ref{Mup}) and 
 and are ($4.18$, $560$, $0.095$, $4.8\times10^{-3}$, $640$) for Eq.(\ref{Mdown}). The masses of the three generation
 of quarks are as given by the PDG~\cite{Olive:2016xmw}.   All masses are in GeV and all phases in rad.}
\label{table:1}
\end{center}
\end{table}

\begin{table}[H]
\begin{center}
\begin{tabular}{l  c  c  c  c c}
\hline\hline
 & $BR(t\to h^0+c)\% $ & $BR(t\to h^0+u)\% $  & $BR(H_1\to s \bar{b} + \bar s b)\% $ & $BR(H_1\to d \bar{b} + \bar d b)\% $ & $M_{H_1}$ \\
\cline{2-5}
 1 & $0.13$ \%      &     $0.015$ \%          &       $0.49$ \%                     & $0.30$ \%         & $124.8$ \\
 2 & $0.22$ \%      &     $0.025$ \%          &       $0.49$ \%                     & $0.31$ \%         & $125.3$ \\
 3 & $0.40$ \%      &     $0.045$ \%          &       $0.49$ \%                     & $0.31$ \%         & $125.4$ \\
 4 & $0.59$ \%      &     $0.066$ \%          &       $0.49$ \%                     & $0.31$ \%         & $125.4$ \\
\hline\hline
\end{tabular}
\caption{An exhibition of the branching ratios for the flavor violating top decays  $t\to c h$ and $t\to h u$ 
and also for the flavor violating Higgs decay $h^0\to \sb$ { and $h^0\to \db$}.
 The results of the table
are consistent with the experimental data of Eq.(1). The analysis is for the parameter sets given by
(1): $\tan\beta=20$, $m_0=m^u_0=m^d_0=5000$, $|\mu|=650$, $|A^u_0|=400$, $|A^d_0|=210$;
 (2): $\tan\beta=30$, $m_0=m^u_0=m^d_0=6400$, $|\mu|=580$, $|A^u_0|=260$, $|A^d_0|=140$;
 (3): $\tan\beta=40$, $m_0=m^u_0=m^d_0=7100$, $|\mu|=490$, $|A^u_0|=170$, $|A^d_0|=150$;
(4): $\tan\beta=50$, $m_0=m^u_0=m^d_0=9400$, $|\mu|=540$, $|A^u_0|=150$, $|A^d_0|=150$.
 The common parameters are $\theta_{\mu}=0.1$, $\alpha_{A^u_0}=0.4$, $\alpha_{A^d_0}=0.02$, $m_A=600$.
The values of the parameters $h_3, h_4, h_5, h_3', h_4',h_5',  h_3'', h_4'',h_5'', h_6, h_7, h_8$ and 
$y_1, y_2, y_3, y_4, y_5, y_1', y_2', y_3', y_4', y_5'$  are given table~\ref{table:1}.
All masses are in GeV and all phases in rad.
}
\label{table:2}
\end{center}
\end{table}

\begin{figure}[H]
\begin{center}
{\rotatebox{0}{\resizebox*{7.3cm}{!}{\includegraphics{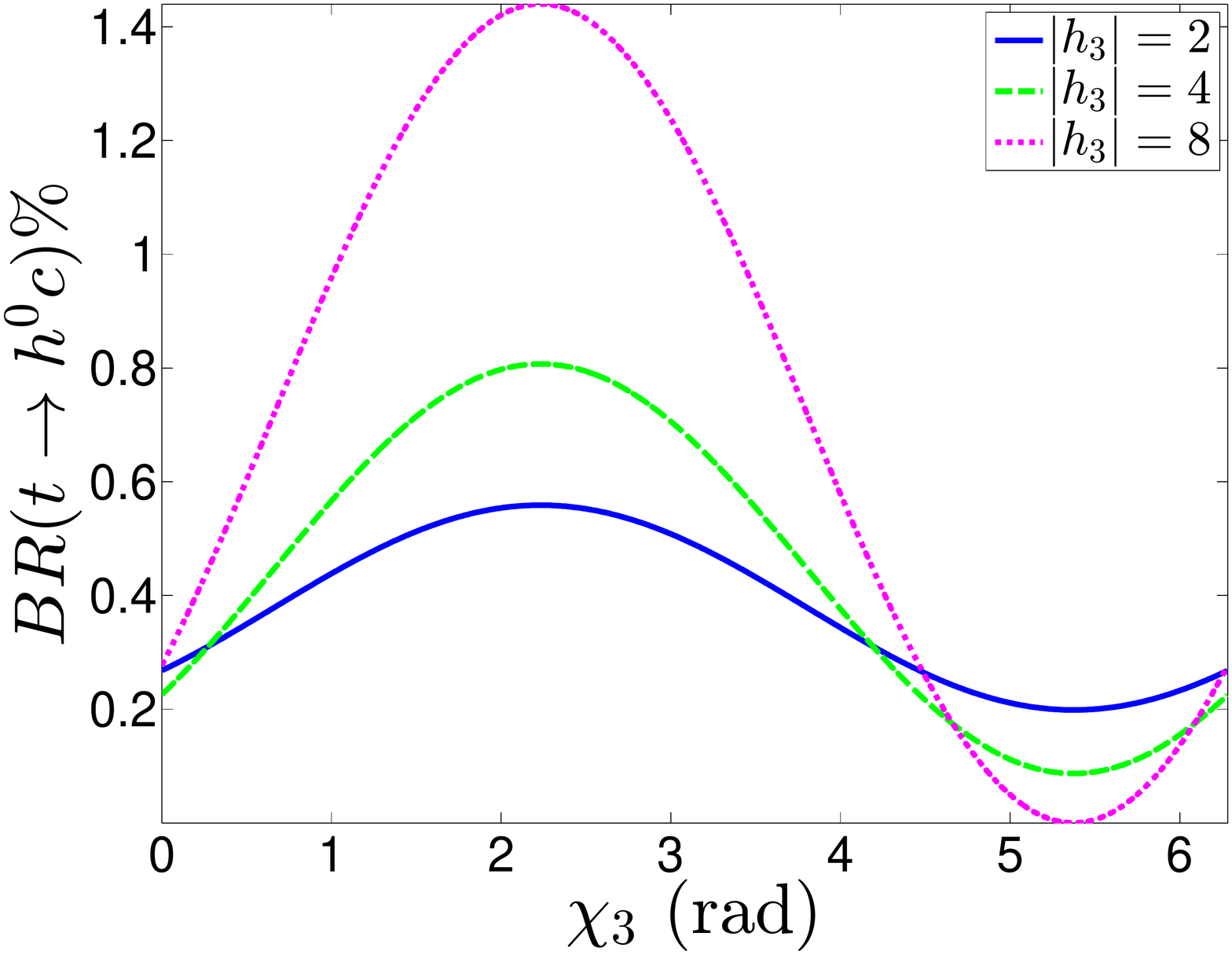}}\hglue5mm}}
{\rotatebox{0}{\resizebox*{7.3cm}{!}{\includegraphics{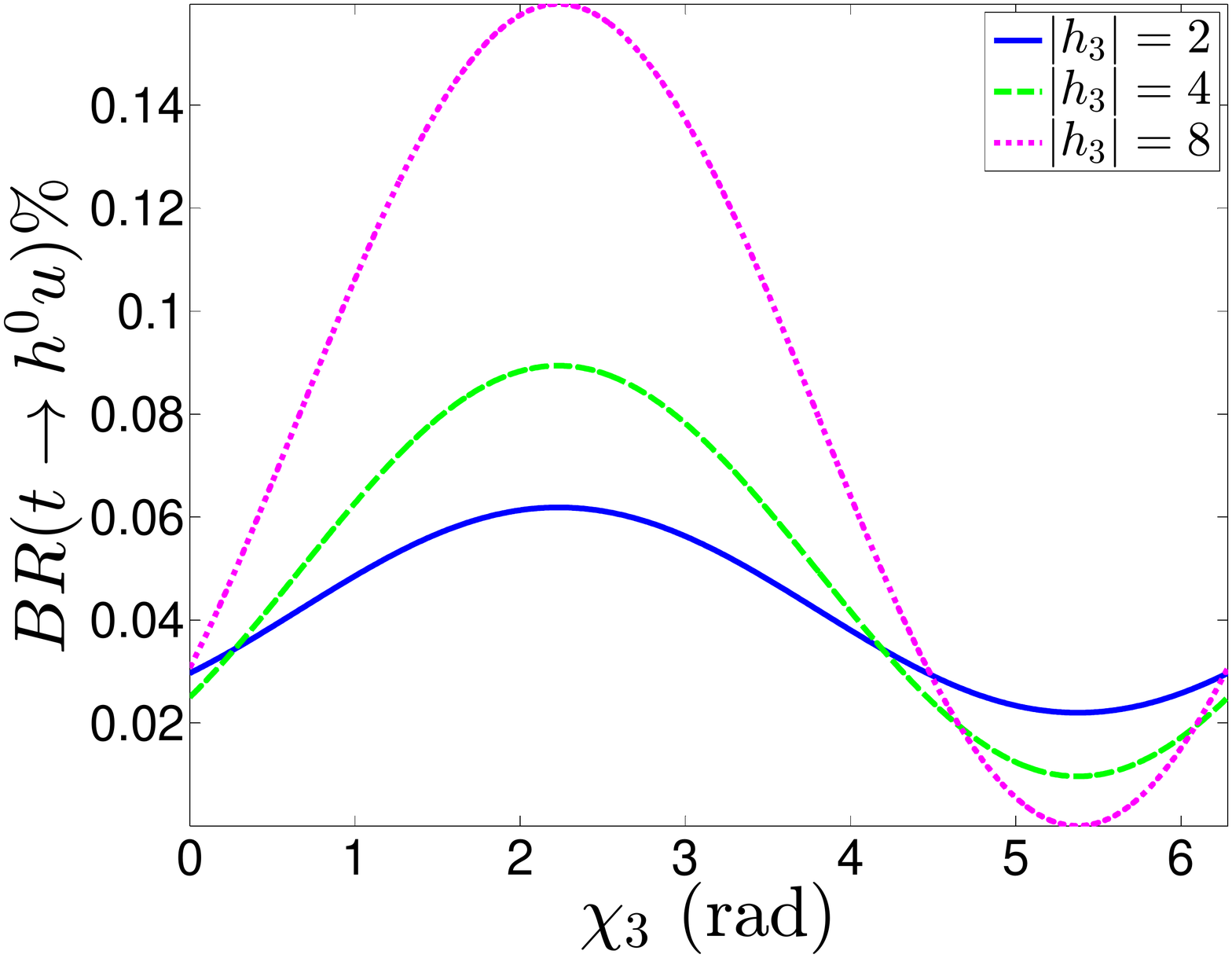}}\hglue5mm}}
\caption{
Left panel: Variation of the $BR(t\to h^0 c)\% $ versus $\chi_3$, for three values of $|h_3|$. From bottom
to top at $\chi_3=1$ (rad), $|h_3|= 2, 4,$ and $8$ GeV.
Other parameters have the values of point 3 in table~\ref{table:1}.
Right panel: Variation of the $BR(t\to h^0 u)\% $ versus $\chi_3$, for three values of $|h_3|$. From bottom
to top at $\chi_3=1$ (rad), $|h_3|= 2, 4,$ and $8$ GeV. 
Other parameters have the values of point 3 in table~\ref{table:1}.
}
\label{fig1}
\end{center}
\end{figure}

\begin{figure}[H]
\begin{center}
{\rotatebox{0}{\resizebox*{7.3cm}{!}{\includegraphics{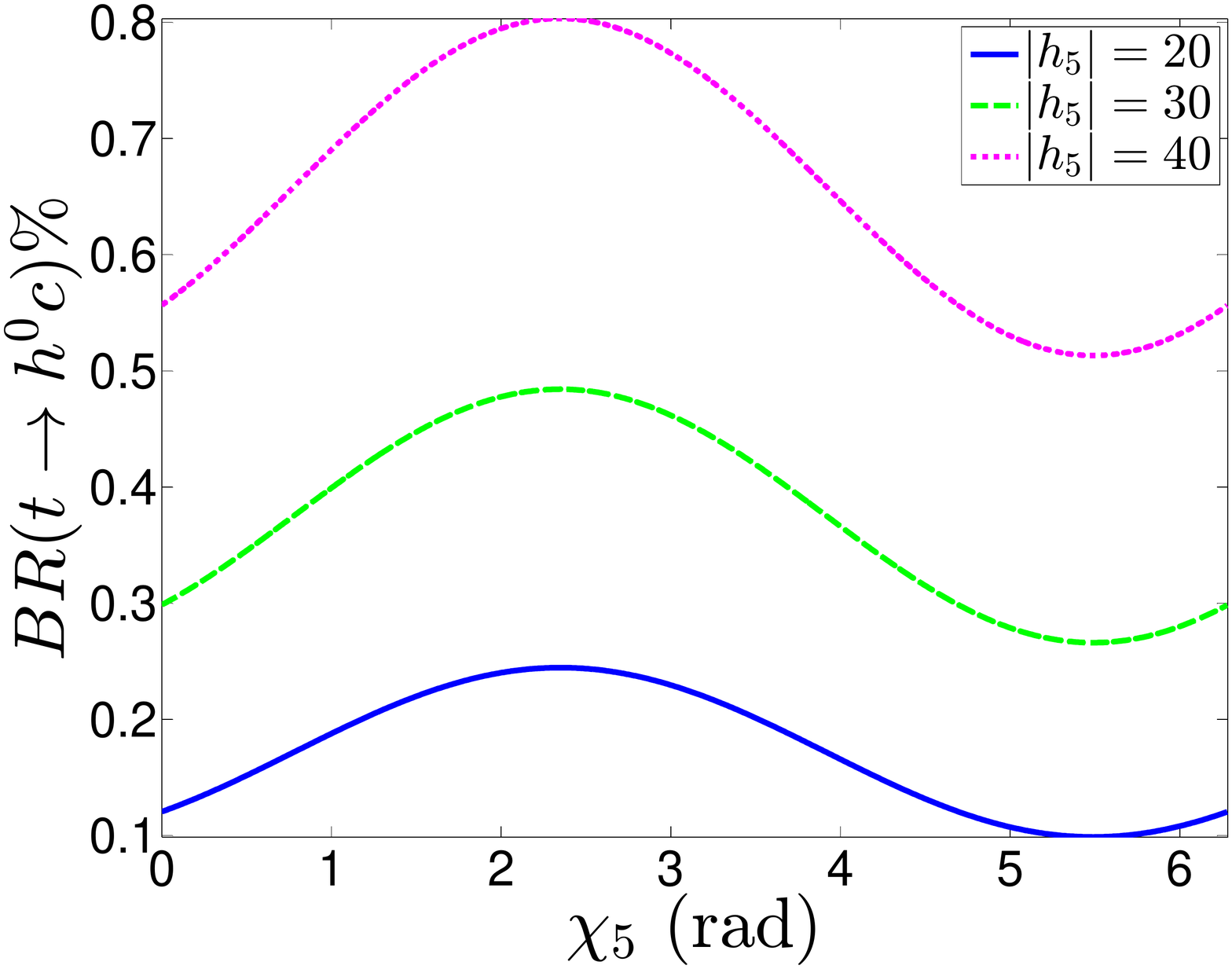}}\hglue5mm}}
{\rotatebox{0}{\resizebox*{7.3cm}{!}{\includegraphics{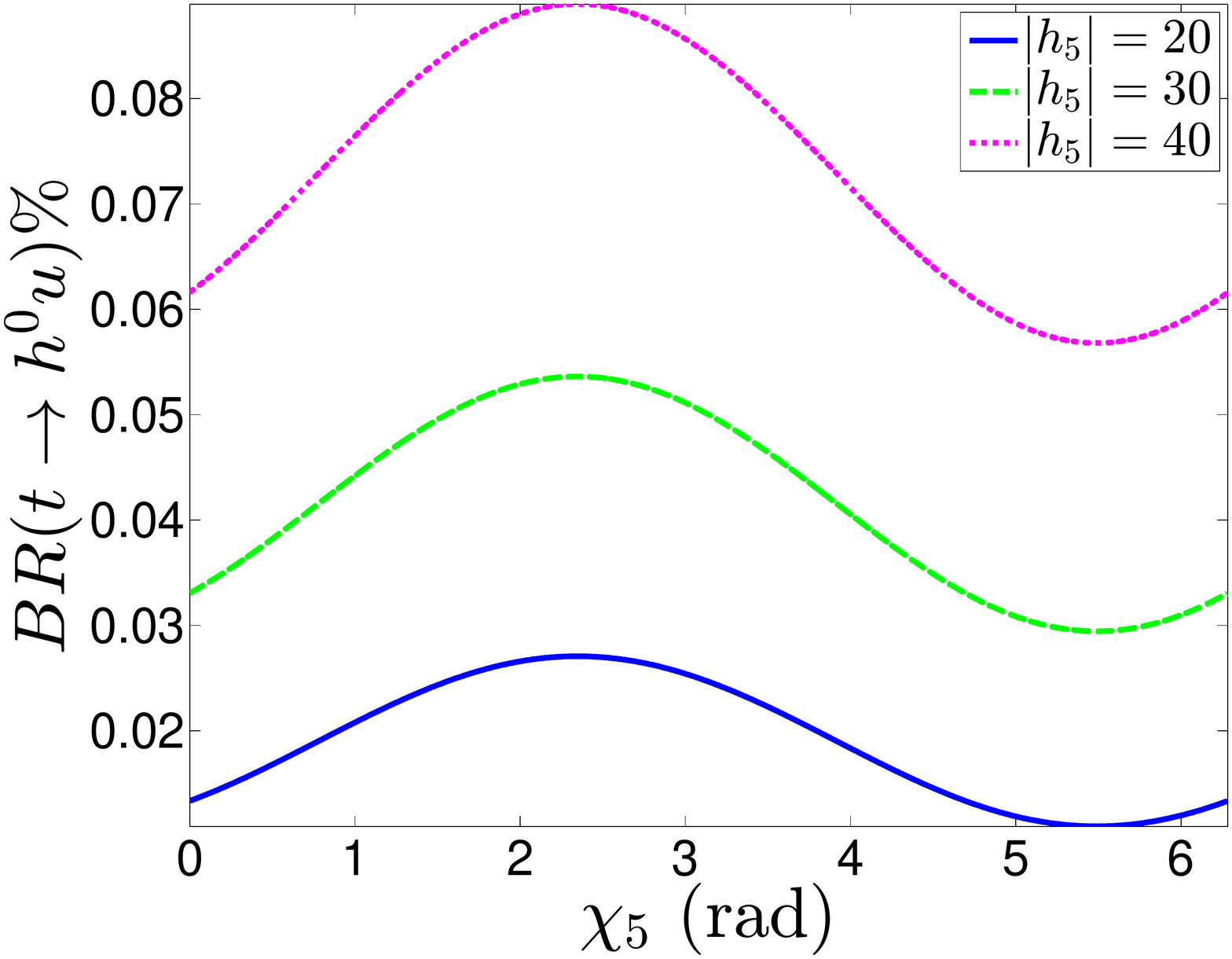}}\hglue5mm}}
\caption{
Left panel: Variation of the $BR(t\to h^0 c)\% $ versus $\chi_5$, for three values of $|h_5|$. From bottom to top $|h_5|= 20, 30,$ and $40$ GeV.
Other parameters have the values of point 2 in table~\ref{table:1}.
Right panel: Variation of the $BR(t\to h^0 u)\% $ versus $\chi_5$, for three values of $|h_5|$. From bottom to top $|h_5|= 20, 30,$ and $40$ GeV.
 Other parameters have the values of point 2 in table~\ref{table:1}.
}
\label{fig2}
\end{center}
\end{figure}

\begin{figure}[H]
\begin{center}
{\rotatebox{0}{\resizebox*{7.3cm}{!}{\includegraphics{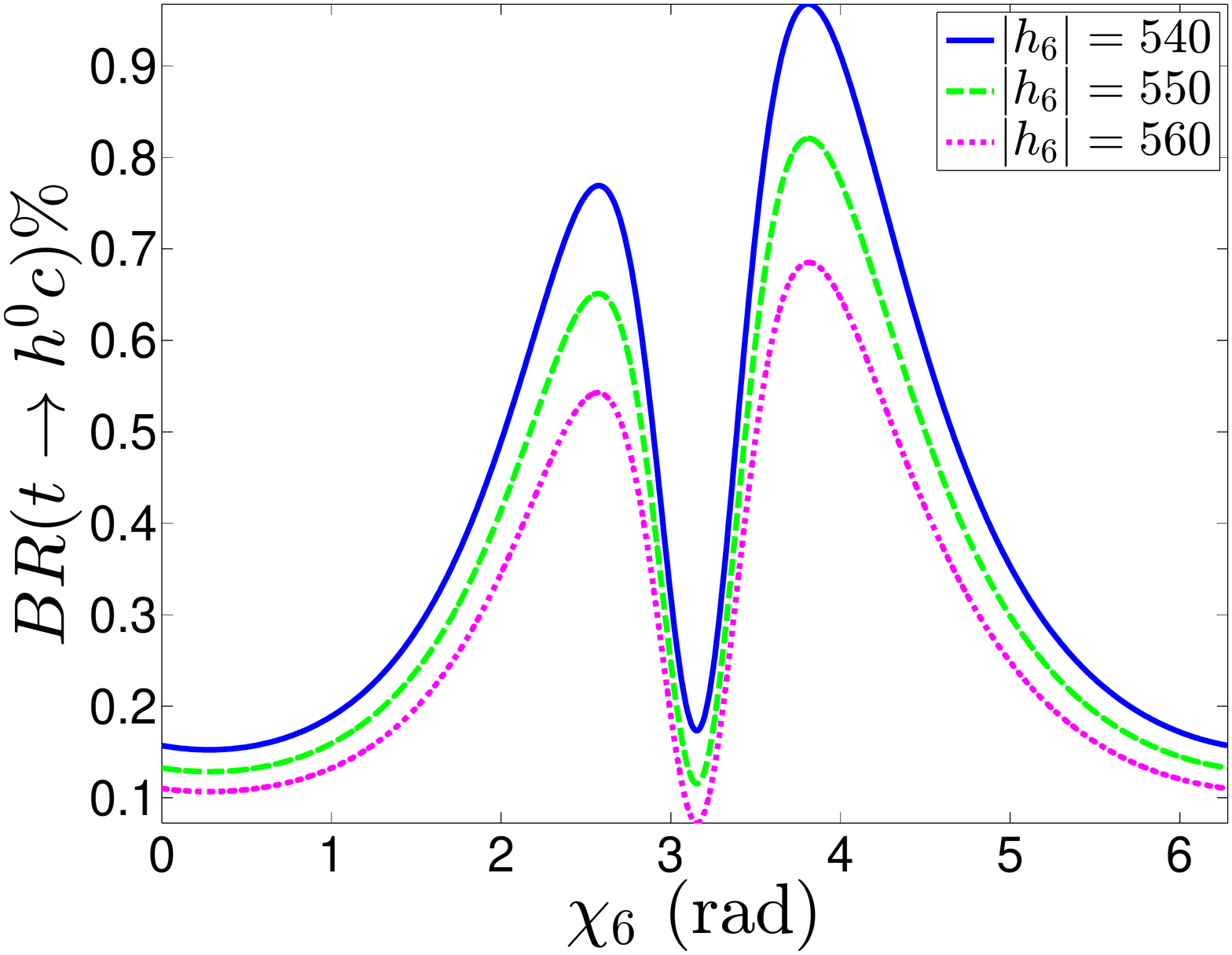}}\hglue5mm}}
{\rotatebox{0}{\resizebox*{7.3cm}{!}{\includegraphics{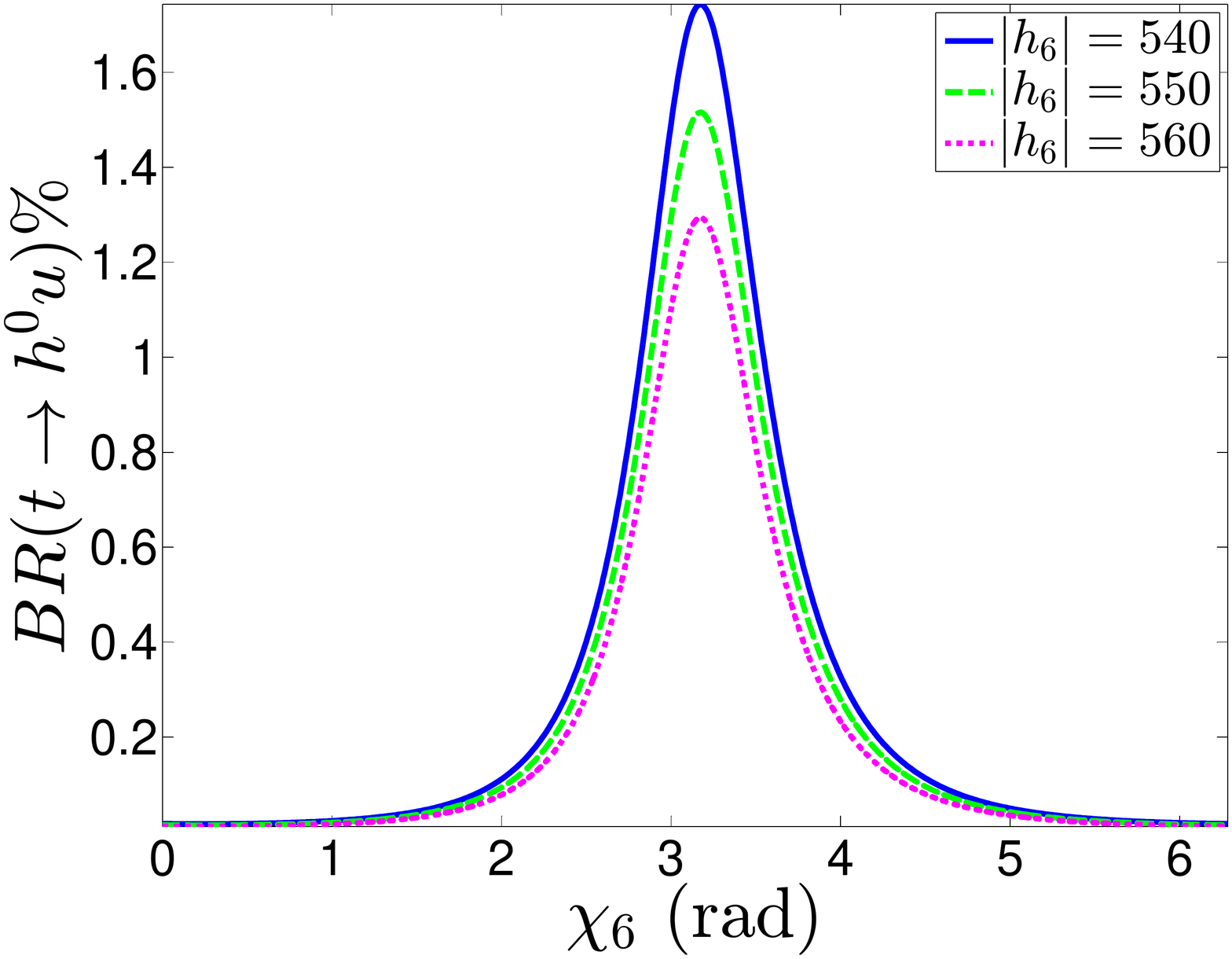}}\hglue5mm}}
\caption{
Left panel: Variation of the $BR(t\to h^0 c)\% $ versus $\chi_6$, for three values of $|h_6|$. From top
to bottom $|h_6|= 540, 550,$ and $560$ GeV. Other parameters have the values of point 1 in table~\ref{table:1}.
Right panel: Variation of the $BR(t\to h^0 u)\% $ versus $\chi_6$, for three values of $|h_6|$. From top
to bottom $|h_6|= 540, 550,$ and $560$ GeV. Other parameters have the values of point 1 in table~\ref{table:1}.
}
\label{fig3}
\end{center}
\end{figure}

\begin{figure}[H]
\begin{center}
{\rotatebox{0}{\resizebox*{7.3cm}{!}{\includegraphics{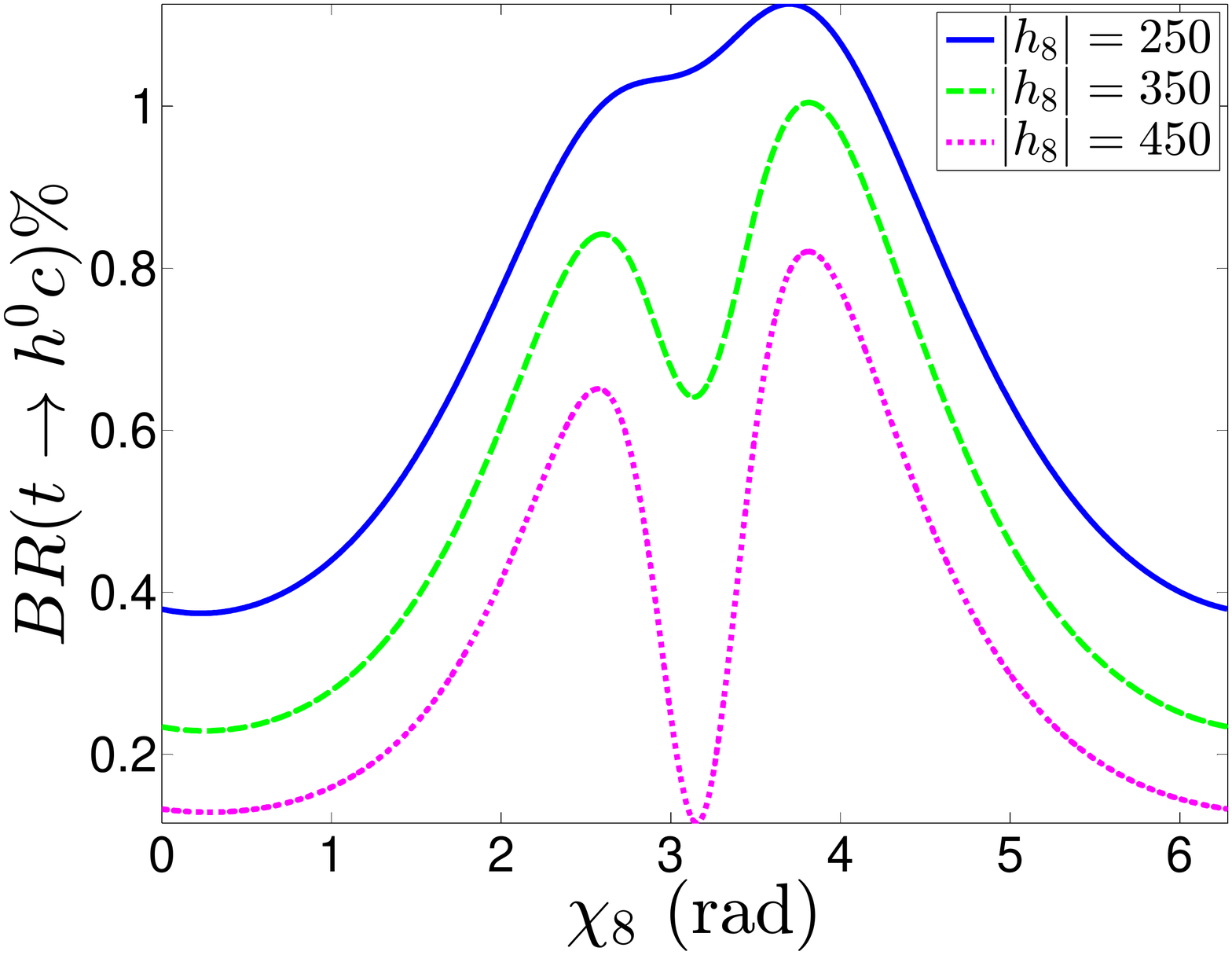}}\hglue5mm}}
{\rotatebox{0}{\resizebox*{7.3cm}{!}{\includegraphics{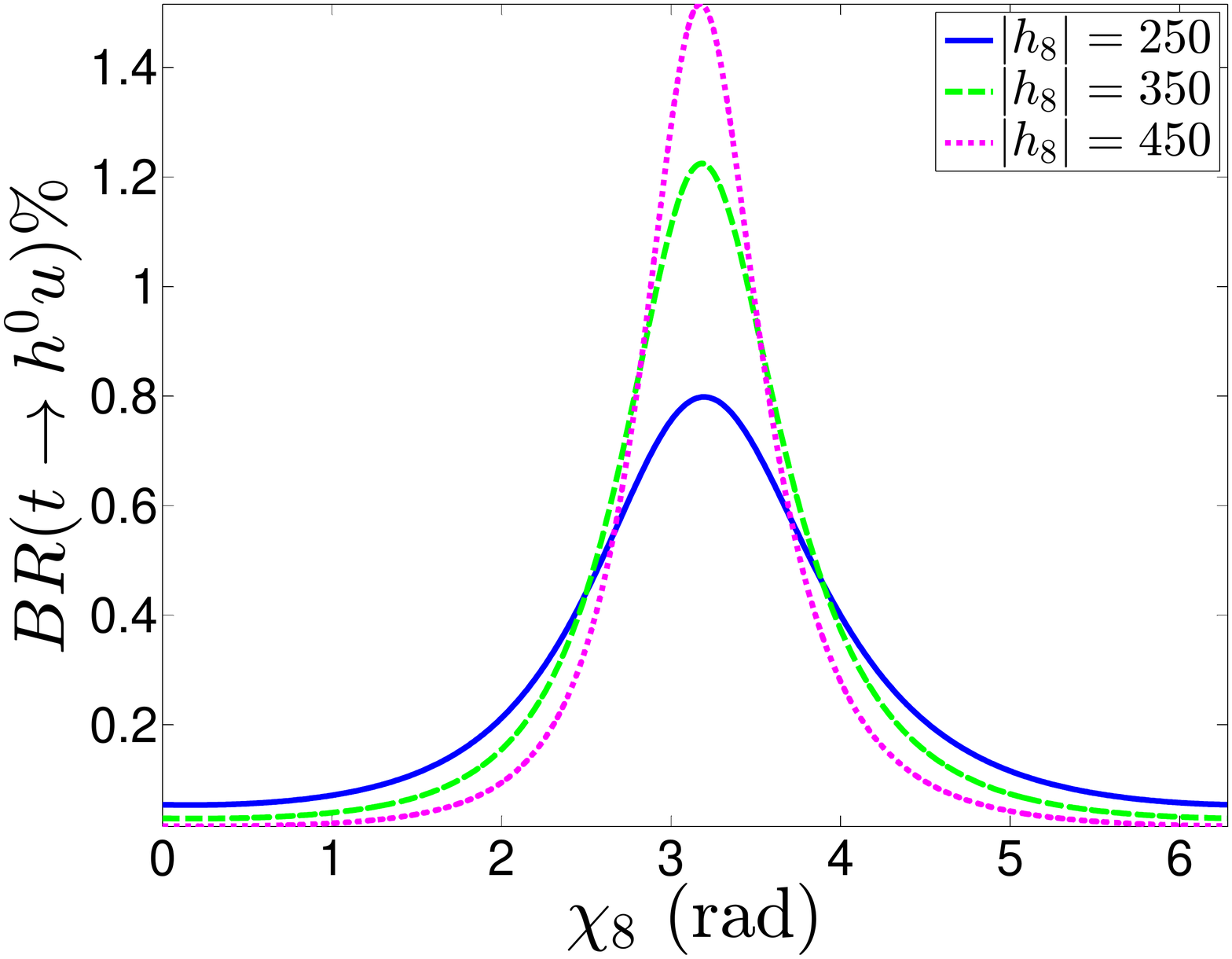}}\hglue5mm}}
\caption{
Left panel: Variation of the $BR(t\to h^0 c)\% $ versus $\chi_8$, for three values of $|h_8|$. From top
to bottom $|h_8|= 250, 350,$ and $450$ GeV. Other parameters have the values of point 1 in table~\ref{table:1}.
Right panel: Variation of the $BR(t\to h^0 u)\% $ versus $\chi_8$, for three values of $|h_8|$. From bottom
to top at $\chi_8=3$ (rad), $|h_8|= 250, 350,$ and $450$ GeV. Other parameters have the values of point 1 in table~\ref{table:1}.
}
\label{fig4}
\end{center}
\end{figure}

\begin{figure}[H]
\begin{center}
{\rotatebox{0}{\resizebox*{7.3cm}{!}{\includegraphics{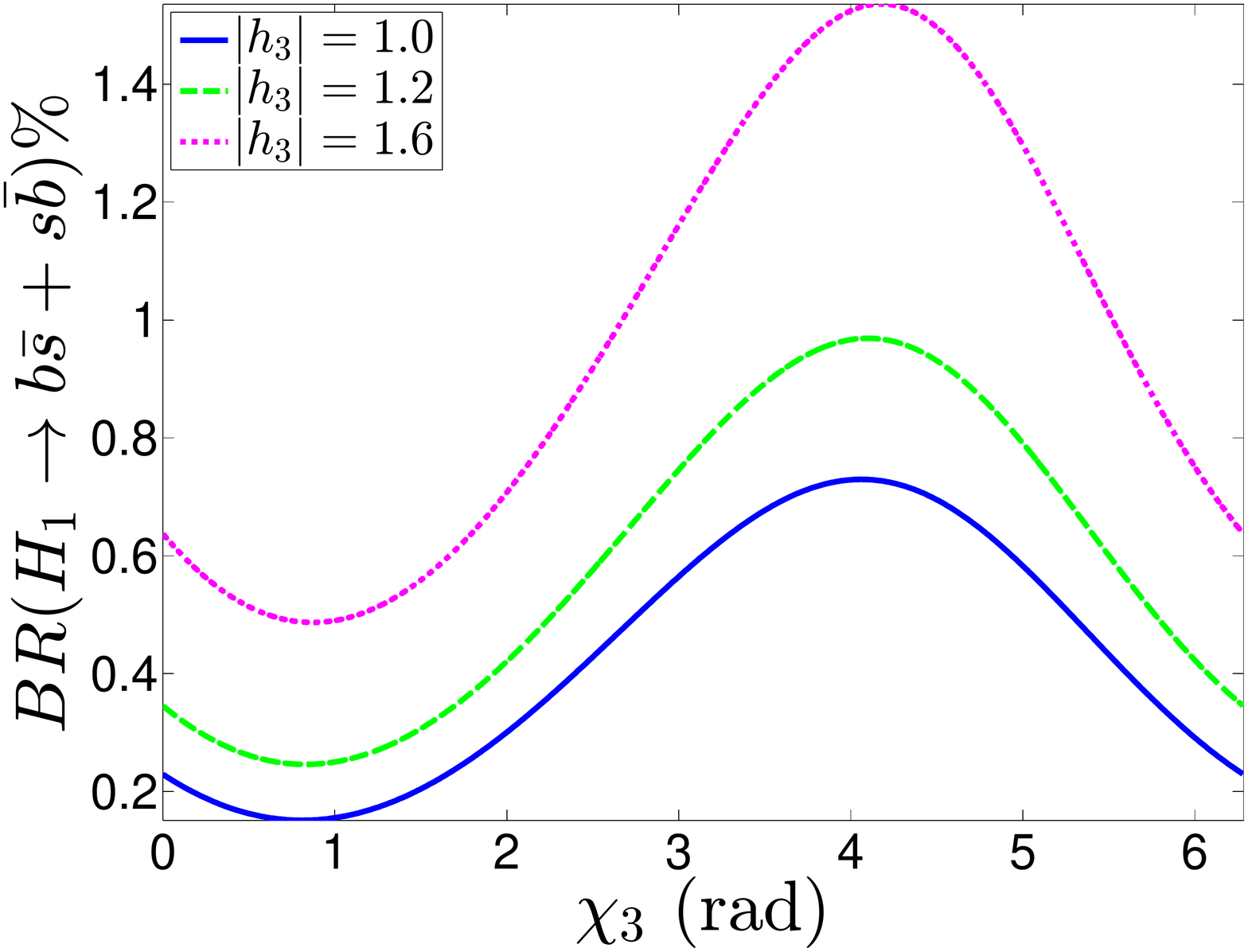}}\hglue5mm}}
{\rotatebox{0}{\resizebox*{7.3cm}{!}{\includegraphics{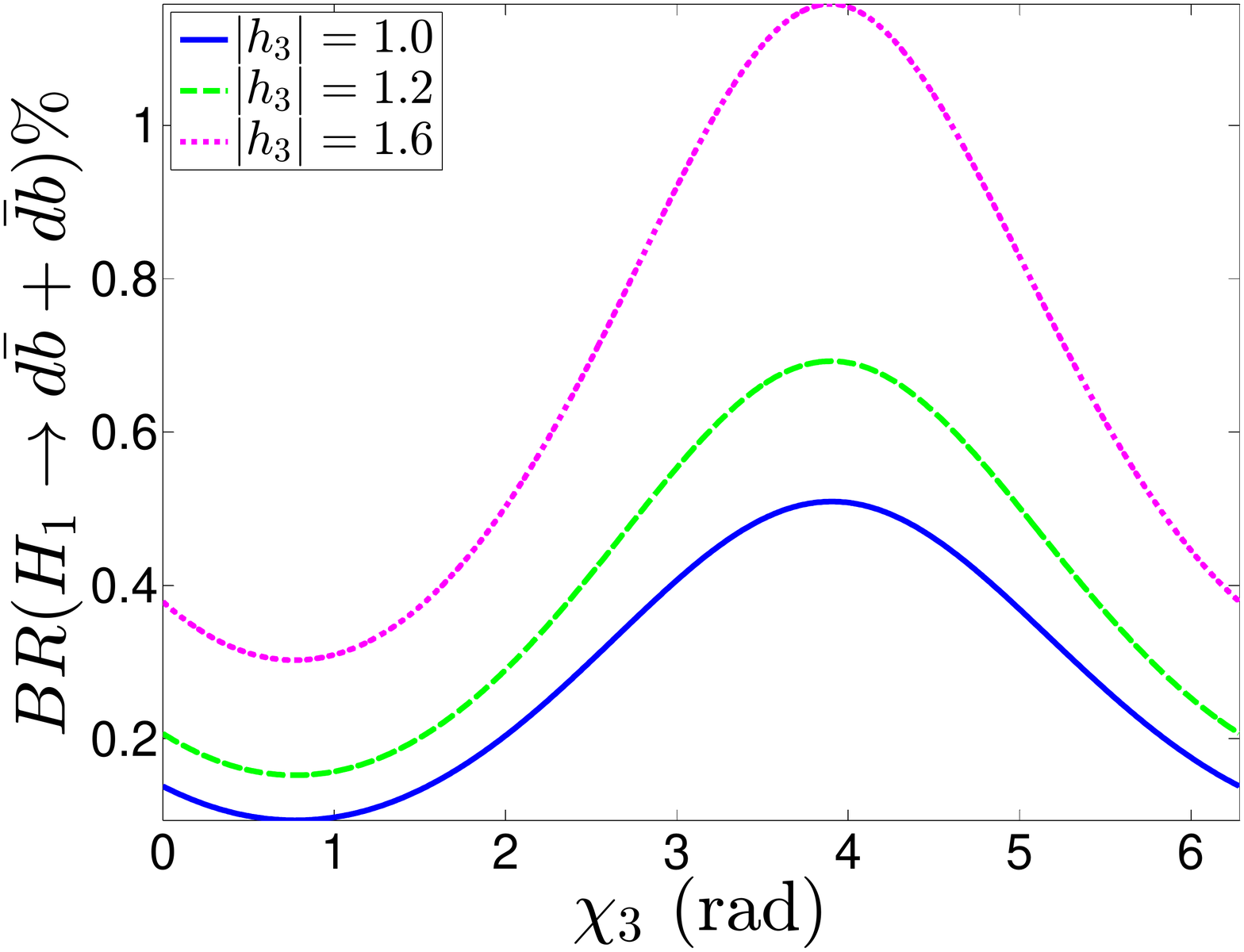}}\hglue5mm}}
\caption{
Left panel: Variation of the $BR(H_1\to b \bar{s} + s \bar{b})\%$ versus $\chi_3$, for three values of $|h_3|$. From bottom
to top  $|h_3|= 1.0, 1.2,$ and $1.6$ GeV. Other parameters have the values of point 1 in table~\ref{table:1}.
Right panel: Variation of the $BR(H_1\to d\bar{b}+ \bar{d}b)\%$ versus $\chi_3$, for three values of $|h_3|$. From bottom
to top $|h_3|= 1.0, 1.2,$ and $1.6$ GeV. Other parameters have the values of point 1 in table~\ref{table:1}.
}
\label{fig5}
\end{center}
\end{figure}

\begin{figure}[H]
\begin{center}
{\rotatebox{0}{\resizebox*{7.3cm}{!}{\includegraphics{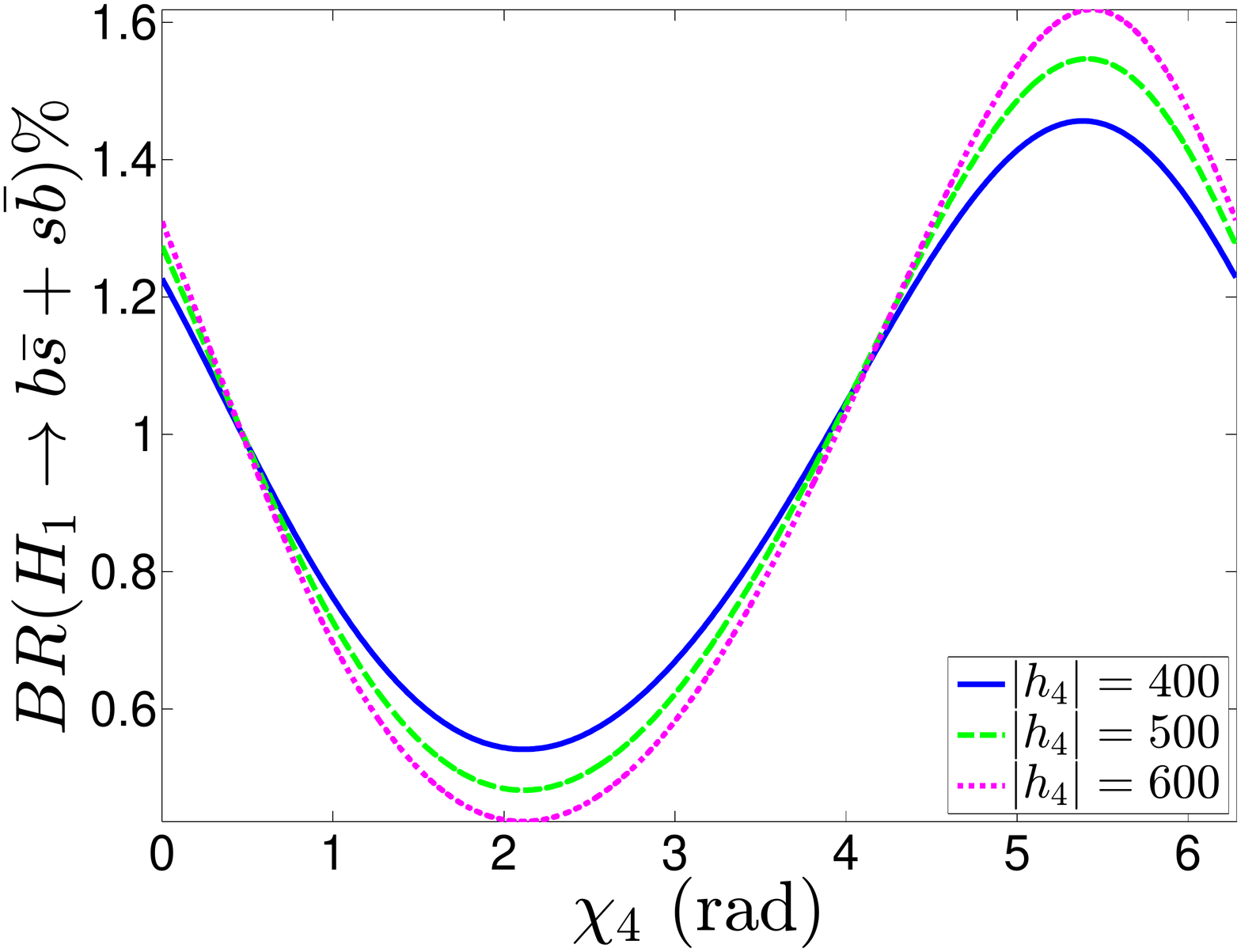}}\hglue5mm}}
{\rotatebox{0}{\resizebox*{7.3cm}{!}{\includegraphics{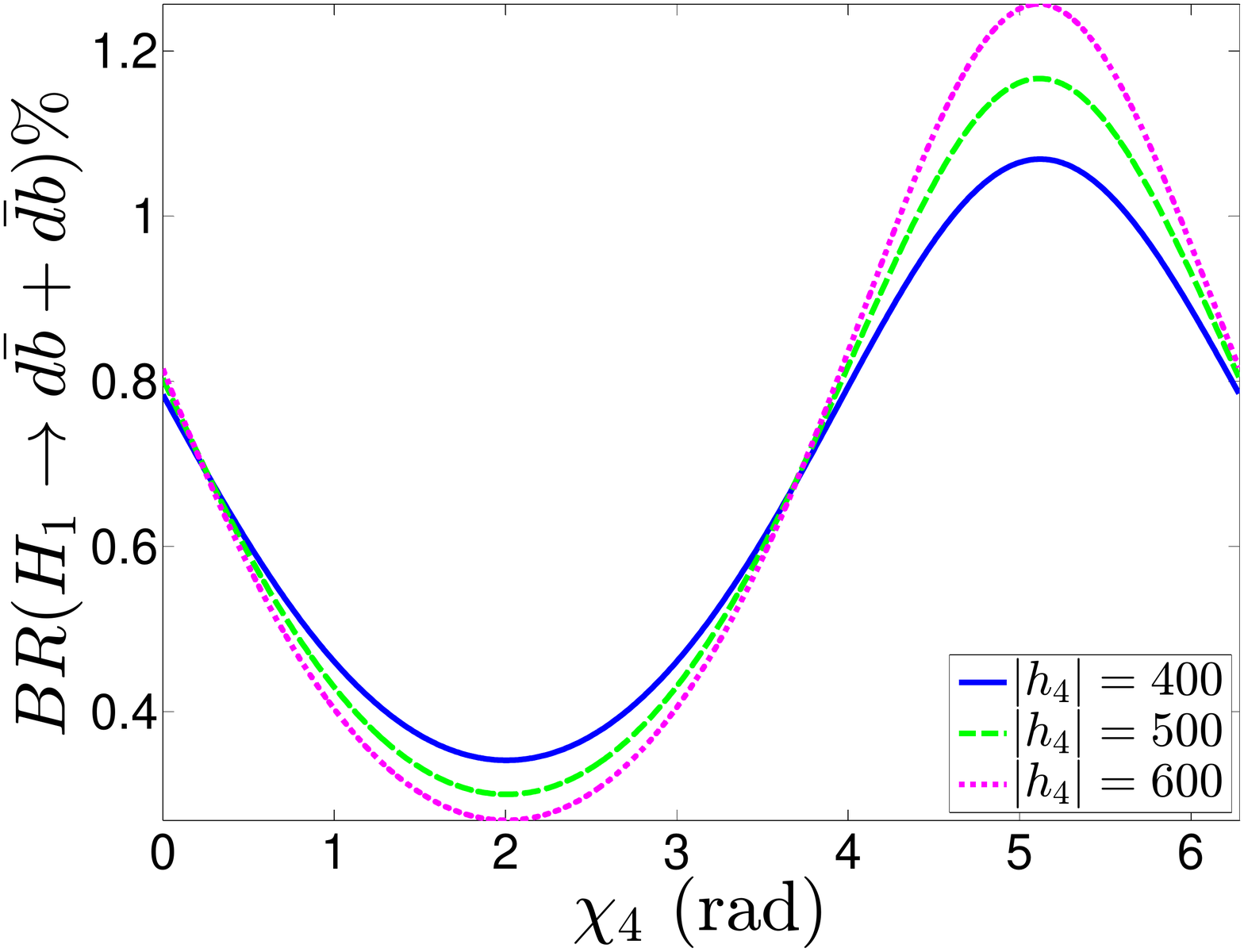}}\hglue5mm}}
\caption{
Left panel: Variation of the $BR(H_1\to b \bar{s} + s \bar{b})\%$ versus $\chi_4$, for three values of $|h_4|$. From bottom
to top at $\chi_4=5$ (rad), $|h_4|= 400, 500,$ and $600$ GeV. Other parameters have the values of point 2 in table~\ref{table:1}.
Right panel: Variation of the $BR(H_1\to d\bar{b}+ \bar{d}b)\%$ versus $\chi_4$, for three values of $|h_4|$. From bottom
to top at $\chi_4=5$ (rad), $|h_4|= 400, 500,$ and $600$ GeV. Other parameters have the values of point 2 in table~\ref{table:1}.
}
\label{fig6}
\end{center}
\end{figure}

\begin{figure}[H]
\begin{center}
{\rotatebox{0}{\resizebox*{7.3cm}{!}{\includegraphics{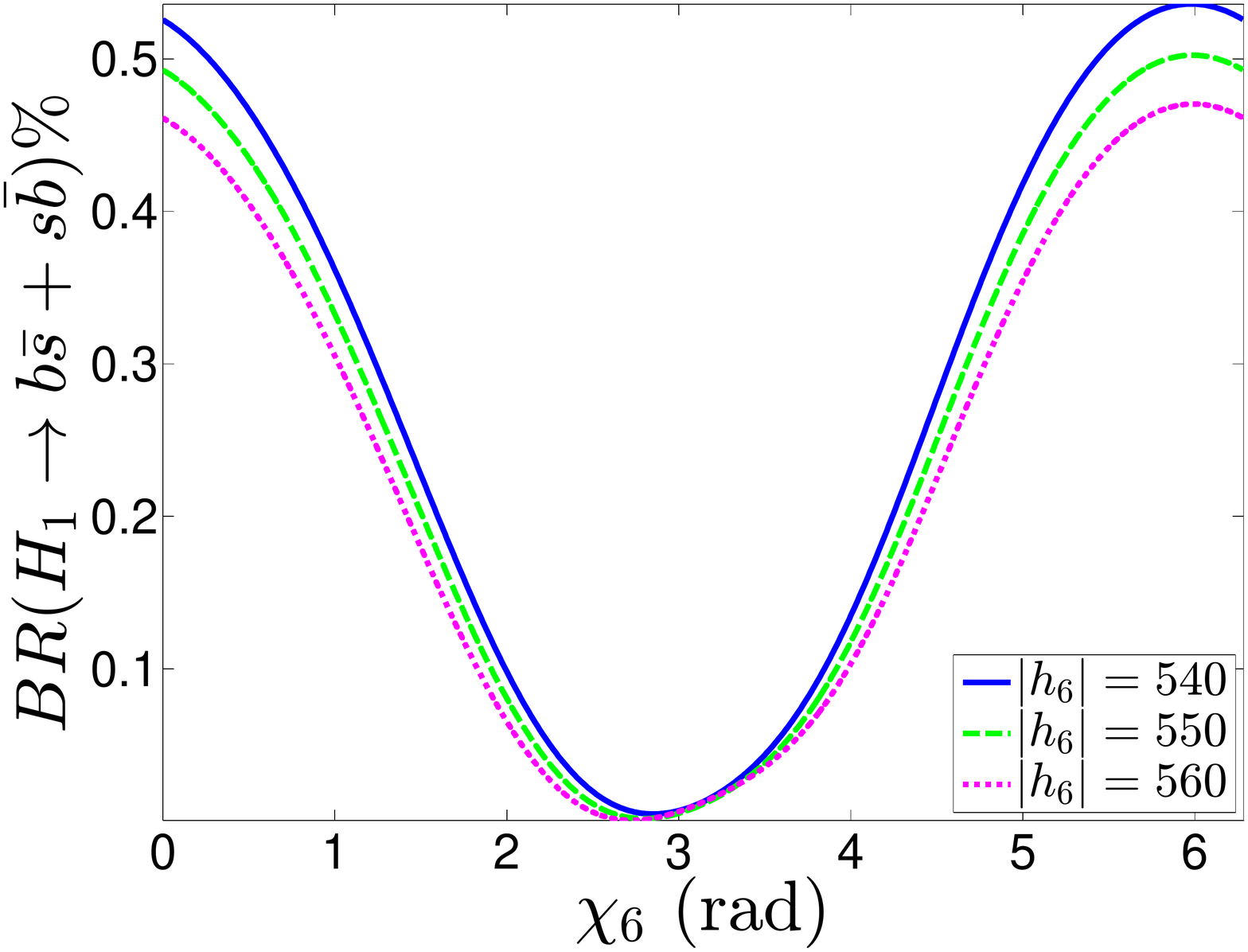}}\hglue5mm}}
{\rotatebox{0}{\resizebox*{7.3cm}{!}{\includegraphics{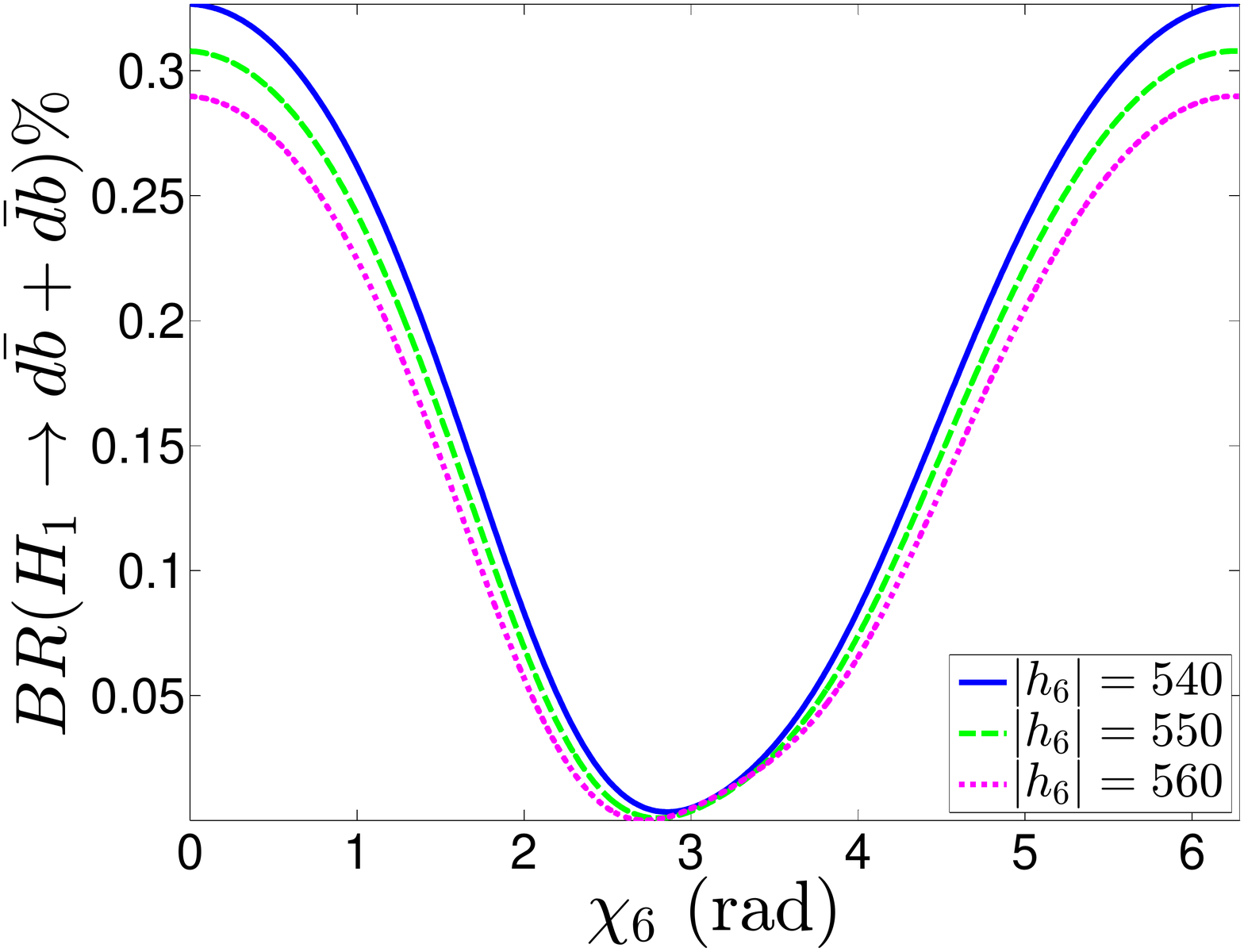}}\hglue5mm}}
\caption{
Left panel: Variation of the $BR(H_1\to b \bar{s} + s \bar{b})\%$ versus $\chi_6$, for three values of $|h_6|$. From top
to bottom $|h_6|= 540, 550,$ and $560$ GeV. Other parameters have the values of point 3 in table~\ref{table:1}.
Right panel: Variation of the $BR(H_1\to d\bar{b}+ \bar{d}b)\%$ versus $\chi_6$, for three values of $|h_6|$. From top
to bottom $|h_6|= 540, 550,$ and $560$ GeV. Other parameters have the values of point 3 in table~\ref{table:1}.
}
\label{fig7}
\end{center}
\end{figure}

\begin{figure}[H]
\begin{center}
{\rotatebox{0}{\resizebox*{7.3cm}{!}{\includegraphics{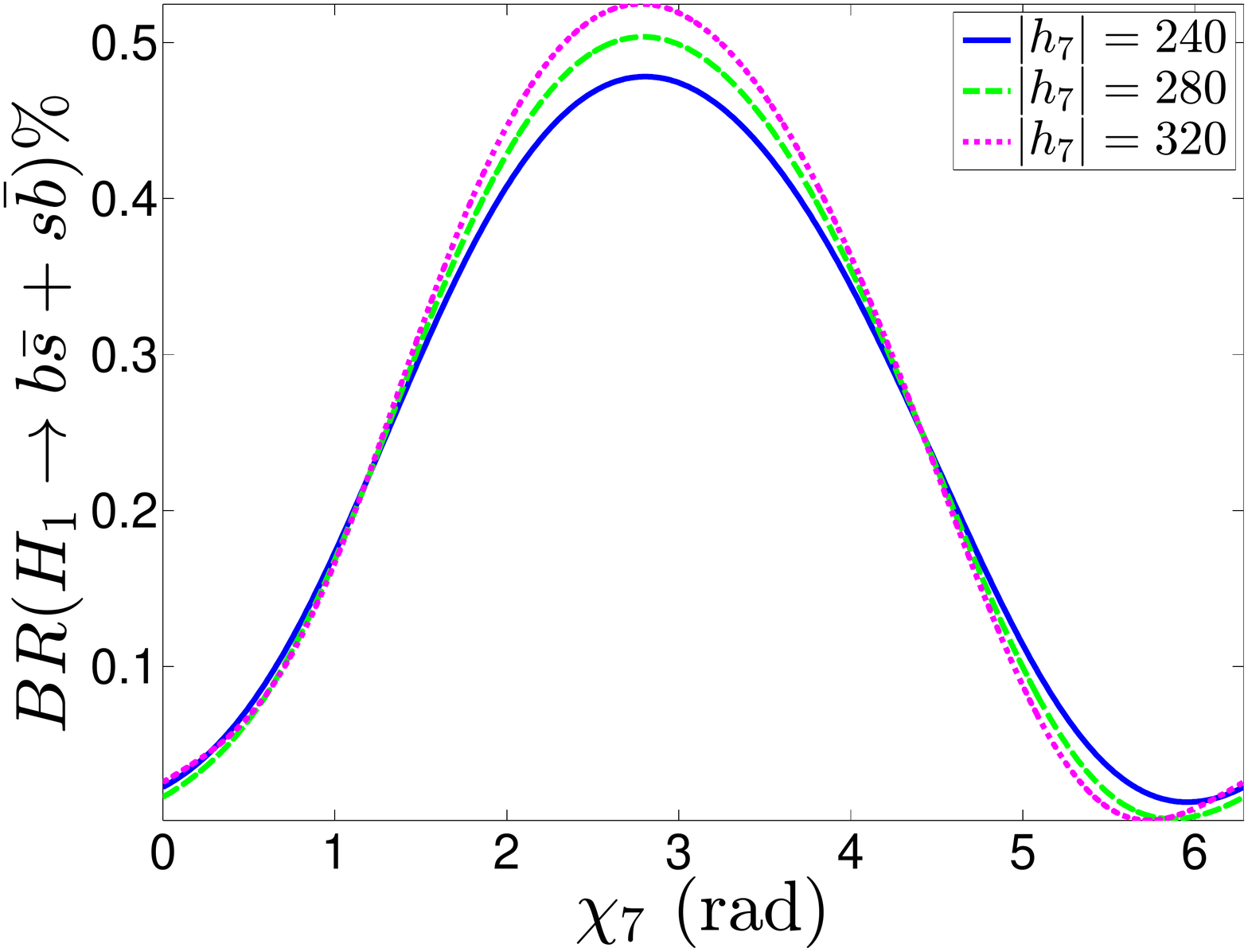}}\hglue5mm}}
{\rotatebox{0}{\resizebox*{7.3cm}{!}{\includegraphics{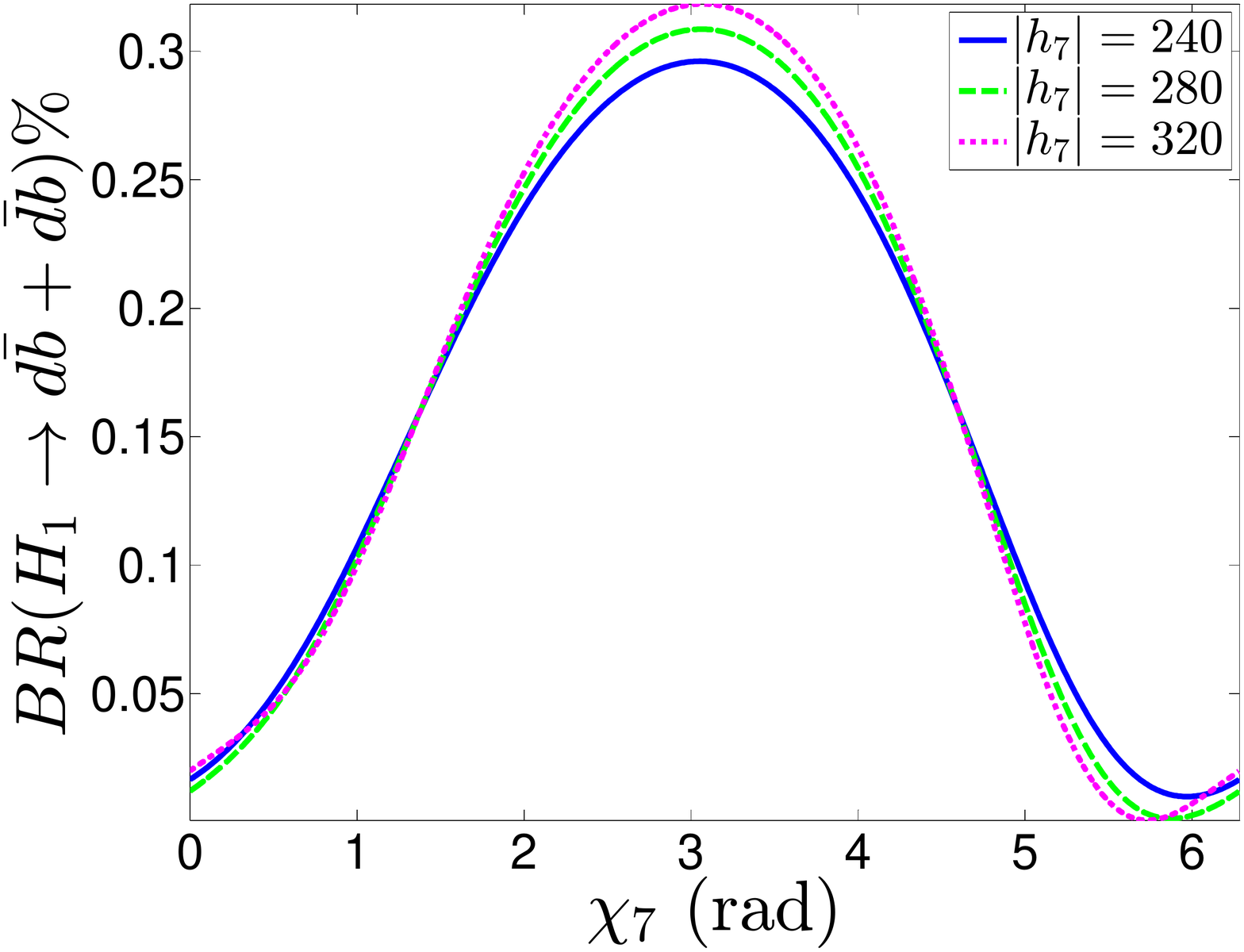}}\hglue5mm}}
\caption{
Left panel: Variation of the $BR(H_1\to b \bar{s} + s \bar{b})\%$ versus $\chi_7$, for three values of $|h_7|$. From bottom
to top at $\chi_7=3$ (rad), $|h_7|= 240, 280,$ and $320$ GeV. Other parameters have the values of point 4 in table~\ref{table:1}.
Right panel: Variation of the $BR(H_1\to d\bar{b}+ \bar{d}b)\%$ versus $\chi_7$, for three values of $|h_7|$. From bottom
to top at $\chi_7=3$ (rad), $|h_7|= 240, 280,$ and $320$ GeV. Other parameters have the values of point 4 in table~\ref{table:1}.
}
\label{fig8}
\end{center}
\end{figure}

    The dependence of the \flv branching ratios on CP phases are discussed in Figs. (\ref{fig1})-(\ref{fig8}).
     In  Fig. (\ref{fig1}) we exhibit the dependence of $t\to h^0 c$ and $t\to h^0 u$ on the CP phase $\chi_3$.
     A similar analysis for these branching ratios on $\chi_5$ is given in Fig.(\ref{fig2}),  on 
     $\chi_6$  in Fig. (\ref{fig3}), and on $\chi_8$  in Fig. (\ref{fig4}). In these 
   analyses one finds that the branching ratios are  sensitive to the phases. 
   A similar analysis for the \flv branching ratio of the Higgs $H_1\to \sb$ { and $H_1\to \db$} are given in 
   Figs. (\ref{fig5})-(\ref{fig8}). In Fig. (\ref{fig5}) the analysis is for the dependence on the CP phase
   $ \chi_3$, { on $\chi_4$ in Fig. (\ref{fig6}), on $\chi_6$ in Fig. (\ref{fig7}), 
   and on $\chi_7$ in Fig. (\ref{fig8})}. As in the case of \flv decays of the top 
   here too we find that the decays are sensitive to the CP phases. We note that in all these
   cases the mixings with the vector like generation enter prominently.


\section{Conclusion\label{sec6}}

 As is well known flavor violating processes involving the Higgs and the quarks  are highly suppressed in the standard 
model and beyond the reach of experimental observation. They are also suppressed in MSSM
and beyond the reach of experiment. 
In this work we have analyzed such processes in the framework of an extended MSSM with a  vector like
quark generation. In this framework we first analyze the flavor violating top decays  $t\to h^0 c$ and $t\to h^0 u$. 
 Here it is shown that  branching ratios can be several orders of magnitude larger than in the standard 
 model on in MSSM and could be as large as the upper limits  given by the ATLAS  and the CMS  
 Collaborations.
 Next we analyze the Higgs boson decay $h^0 \to \sb$  and $h^0\to \db$. As in the \flv decays of the top, here also we find that 
 the   branching ratios in this model can be several orders of magnitude larger than in the standard model
 or in MSSM could be as large as $O(1)\%$. Such a branching ratio may be testable with more data 
 from LHC13 and may also lie  within reach of future  colliders specifically the Higgs factories.\\

\textbf{Acknowledgments: }
This research was supported in part by the NSF Grant d PHY-1620575.\\

\section{Appendix A: Squark mass matrices  \label{sec7}}

In this appendix we give further details of the extended MSSM 
model with vector like generation.  As discussed in
section \ref{sec2} we allow for mixing between the vector generation and specifically the mirrors  and the standard three generations of quarks. The superpotential allowing such mixings is given by

\begin{align}
W&=\epsilon_{ij}  [y_{1}  \hat H_1^{i} \hat q_{1L} ^{j}\hat b^c_{1L}
 +y_{1}'  \hat H_2^{j}  \hat q_{1L} ^{i}\hat t^c_{1L}
+y_{2}  \hat H_1^{i} \hat Q^c{^{j}}\hat T_{L}
+y_{2}'  \hat H_2^{j} \hat Q^c{^{i}}\hat B_{L}\nonumber \\
 &+y_{3}  \hat H_1^{i} \hat q_{2L} ^{j}\hat b^c_{2L}
 +y_{3}'  \hat H_2^{j}  \hat q_{2L} ^{i}\hat t^c_{2L}
 +y_{4}  \hat H_1^{i} \hat q_{3L} ^{j}\hat b^c_{3L}
 +y_{4}'  \hat H_2^{j}  \hat q_{3L} ^{i}\hat t^c_{3L}
+y_{5}  \hat H_1^{i} \hat q_{4L} ^{j}\hat b^c_{4L}
+y_{5}'  \hat H_2^{j}  \hat q_{4L} ^{i}\hat t^c_{4L}
] \nonumber \\
&+ h_{3} \epsilon_{ij}  \hat Q^c{^{i}}\hat q_{1L}^{j}+
h_{3}' \epsilon_{ij}  \hat Q^c{^{i}}\hat q_{2L}^{j}+
h_{3}'' \epsilon_{ij}  \hat Q^c{^{i}}\hat q_{3L}^{j}+
h_{6} \epsilon_{ij}  \hat Q^c{^{i}}\hat q_{4L}^{j}
+h_4 \hat b_{1L}^c \hat B_{L} +h_5 \hat t_{1L}^c \hat T_{L}\nonumber\\
&+h_4'  \hat b_{2L}^c \hat B_{L} +h_5' \hat t_{2L}^c \hat T_{L}
+h_4''  \hat b_{3L}^c \hat B_{L} +h_5'' \hat t_{3L}^c \hat T_{L}
+h_7  \hat b_{4L}^c \hat B_{L} +h_8 \hat t_{4L}^c \hat T_{L}
  -\mu \epsilon_{ij} \hat H_1^i \hat H_2^j \,.
 \label{7w}
\end{align}
Here the couplings are in general complex. Thus, for example,
 $\mu$ is the complex Higgs mixing parameter so that $\mu= |\mu| e^{i\theta_\mu}$.
The mass terms for the 
up quarks (ups), the mirror up quarks, the down quarks, and the mirror down quarks
arise from the term
\beq
{\cal{L}}=-\frac{1}{2}\frac{\partial ^2 W}{\partial{A_i}\partial{A_j}}\psi_ i \psi_ j+\text{h.c.},
\label{6}
\eeq
where $\psi$ and $A$ stand for generic two-component fermion and scalar fields.
After spontaneous breaking of the electroweak symmetry, ($\langle H_1^1 \rangle=v_1/\sqrt{2} $ and $\langle H_2^2\rangle=v_2/\sqrt{2}$),
we have the following set of mass terms written in the four-component spinor notation
so that
\beq
-{\cal L}_m= \bar\xi_R^T (M_u) \xi_L +\bar\eta_R^T(M_{d}) \eta_L +\text{h.c.},
\eeq
where the basis vectors are defined in  Eq. (5) and Eq. (10).

Next we consider the mixings among the squarks and mirror squarks. The  interactions that contribute to them
receive $F$ type and $D$ type contributions as well contributions from soft terms. Thus the terms that 
contribute to the mixings are given by

\beq
{\cal L}= {\cal L}_F +{\cal L}_D + {\cal L}_{\rm soft}\ ,
\label{mass.1}
\eeq
where $- {\cal L}_F=V_F=F_i F^{*}_i$ and   $F_i =\partial W/\partial A_i$ while  ${\cal L}_D$ is given by
\begin{align}
-{\cal L}_D&=\frac{1}{2} m^2_Z \cos^2\theta_W \cos 2\beta \{\tilde t_{ L} \tilde t^*_{ L} -\tilde b_L \tilde b^*_L
+\tilde c_{ L} \tilde c^*_{ L} -\tilde s_L \tilde s^*_L
+\tilde u_{ L} \tilde u^*_{ L} -\tilde d_L \tilde d^*_L
+\tilde t_{4 L} \tilde t^*_{4 L} -\tilde b_{4L} \tilde b^*_{4L}
 \nonumber \\
&+\tilde B_R \tilde B^*_R -\tilde T_R \tilde T^*_R\}
+\frac{1}{2} m^2_Z \sin^2\theta_W \cos 2\beta \{-\frac{1}{3}\tilde t_{ L} \tilde t^*_{ L}
 +\frac{4}{3}\tilde t_{ R} \tilde t^*_{ R}
-\frac{1}{3}\tilde c_{ L} \tilde c^*_{ L}
 +\frac{4}{3}\tilde c_{ R} \tilde c^*_{ R} \nonumber \\
&-\frac{1}{3}\tilde u_{ L} \tilde u^*_{ L}
 +\frac{4}{3}\tilde u_{ R} \tilde u^*_{ R}
+\frac{1}{3}\tilde T_{ R} \tilde T^*_{ R}
 -\frac{4}{3}\tilde T_{ L} \tilde T^*_{ L}
-\frac{1}{3}\tilde b_{ L} \tilde b^*_{ L}
 -\frac{2}{3}\tilde b_{ R} \tilde b^*_{ R}\nonumber\\
&-\frac{1}{3}\tilde s_{ L} \tilde s^*_{ L}
 -\frac{2}{3}\tilde s_{ R} \tilde s^*_{ R}
-\frac{1}{3}\tilde d_{ L} \tilde d^*_{ L}
 -\frac{2}{3}\tilde d_{ R} \tilde d^*_{ R}
+\frac{1}{3}\tilde B_{ R} \tilde B^*_{ R}\nonumber\\
&+\frac{2}{3}\tilde B_{ L} \tilde B^*_{ L}
-\frac{1}{3}\tilde t_{4 L} \tilde t^*_{4 L}
 +\frac{4}{3}\tilde t_{ 4R} \tilde t^*_{4 R}
-\frac{1}{3}\tilde b_{4 L} \tilde b^*_{4 L}
 -\frac{2}{3}\tilde b_{ 4R} \tilde b^*_{ 4R}
\}.
\label{12}
\end{align}
For ${\cal L}_{\rm soft}$ we assume the following form
\begin{align}
-{\cal L}_{\text{soft}}&= M^2_{\tilde 1 L} \tilde q^{k*}_{1 L} \tilde q^k_{1 L}
+ M^2_{\tilde 4 L} \tilde q^{k*}_{4 L} \tilde q^k_{4 L}
+ M^2_{\tilde 2 L} \tilde q^{k*}_{2 L} \tilde q^k_{2 L}
+ M^2_{\tilde 3 L} \tilde q^{k*}_{3 L} \tilde q^k_{3 L}
+ M^2_{\tilde Q} \tilde Q^{ck*} \tilde Q^{ck}
 + M^2_{\tilde t_1} \tilde t^{c*}_{1 L} \tilde t^c_{1 L} \nonumber \\
& + M^2_{\tilde b_1} \tilde b^{c*}_{1 L} \tilde b^c_{1 L}
+ M^2_{\tilde t_2} \tilde t^{c*}_{2 L} \tilde t^c_{2 L}
+ M^2_{\tilde b_4} \tilde b^{c*}_{4 L} \tilde b^c_{4 L}
+ M^2_{\tilde t_4} \tilde t^{c*}_{4 L} \tilde t^c_{4 L}\nonumber\\
&+ M^2_{\tilde t_3} \tilde t^{c*}_{3 L} \tilde t^c_{3 L}
+ M^2_{\tilde b_2} \tilde b^{c*}_{2 L} \tilde b^c_{2 L}
+ M^2_{\tilde b_3} \tilde b^{c*}_{3 L} \tilde b^c_{3 L}
+ M^2_{\tilde B} \tilde B^*_L \tilde B_L
 +  M^2_{\tilde T} \tilde T^*_L \tilde T_L \nonumber \\
&+\epsilon_{ij} \{y_1 A_{b} H^i_1 \tilde q^j_{1 L} \tilde b^c_{1L}
-y_1' A_{t} H^i_2 \tilde q^j_{1 L} \tilde t^c_{1L}
+y_5 A_{b_4} H^i_1 \tilde q^j_{4 L} \tilde b^c_{4L}
-y_5' A_{t_4} H^i_2 \tilde q^j_{4 L} \tilde t^c_{4L}
+y_3 A_{s} H^i_1 \tilde q^j_{2 L} \tilde b^c_{2L}\nonumber\\
&-y_3' A_{c} H^i_2 \tilde q^j_{2 L} \tilde t^c_{2L}
+y_4 A_{d} H^i_1 \tilde q^j_{3 L} \tilde b^c_{3L}
-y_4' A_{u} H^i_2 \tilde q^j_{3 L} \tilde t^c_{3L}
+y_2 A_{T} H^i_1 \tilde Q^{cj} \tilde T_{L}
-y_2' A_{B} H^i_2 \tilde Q^{cj} \tilde B_{L}
+\text{h.c.}\}\ .
\label{soft}
\end{align}
Here $M_{\tilde 1 L},  M_{\tilde T}$, etc are the soft masses and $A_t, A_{b}$, etc are the trilinear couplings.
The trilinear couplings are in general complex  and we define their phases so that
\begin{gather}
A_b= |A_b| e^{i \alpha_{A_b}} \  ,
 ~~A_{t}=  |A_{t}|
 e^{i\alpha_{A_{t}}} \ ,
  \cdots \ .
\end{gather}
After spontaneous breaking of the electroweak symmetry, when the Higgs bosons develop VEVs, 
 we construct the scalar mass squared matrices using Eqs. (26), (27), and (28).
 Thus  we denote the mass squared matrix for the 
 down squarks and the down mirror squarks  in the basis $(\tilde  b_L, \tilde B_L, \tilde b_R,
\tilde B_R, \tilde s_L, \tilde s_R, \tilde d_L, \tilde d_R,$ 
$\tilde b_{4L}, \tilde b_{4R})$ by $(M^2_{\tilde d})_{ij}= M^2_{ij}$.
This mass squared matrix is  hermitian and can be  diagonalized  by the unitary transformation
\begin{gather}
 \tilde D^{d \dagger} M^2_{\tilde d} \tilde D^{d} = diag (M^2_{\tilde d_1},
M^2_{\tilde d_2}, M^2_{\tilde d_3},  M^2_{\tilde d_4},  M^2_{\tilde d_5},  M^2_{\tilde d_6},  M^2_{\tilde d_7},  M^2_{\tilde d_8}
 M^2_{\tilde d_9},  M^2_{\tilde d_{10}}
 )\ .
\end{gather}
Similarly we write the   mass squared   matrix in the up squark and up mirror squark
 sector in the basis $(\tilde  t_{ L}, \tilde T_L,$
$ \tilde t_{ R}, \tilde T_R, \tilde  c_{ L},\tilde c_{ R}, \tilde u_{ L}, \tilde u_{R}
,\tilde t_{4 L}, \tilde t_{4R}
 )$ and denote it by 
$(M^2_{\tilde u})_{ij}=m^2_{ij}$ which is also a hermitian matrix, and can be 
diagonalized by the  unitary transformation
\begin{equation}
 \tilde D^{u\dagger} M^2_{\tilde u} \tilde D^{u} = \text{diag} (M^2_{\tilde u_1}, M^2_{\tilde u_2}, M^2_{\tilde u_3},  M^2_{\tilde u_4},M^2_{\tilde u_5},  M^2_{\tilde u_6}, M^2_{\tilde u_7}, M^2_{\tilde u_8}
, M^2_{\tilde u_9}, M^2_{\tilde u_{10}}
)\ .
\end{equation}

We display now the matrix elements $M^2_{ij}$ and $m^2_{ij}$. First for $M^2_{ij}$ we have 

\begin{align}
M^2_{11}&= M^2_{\tilde 1 L}+\frac{v^2_1|y_1|^2}{2} +|h_3|^2 -m^2_Z \cos 2 \beta \left(\frac{1}{2}-\frac{1}{3}\sin^2\theta_W\right), \nonumber\\
M^2_{22}&=M^2_{\tilde B}+\frac{v^2_2|y'_2|^2}{2}+|h_4|^2 +|h'_4|^2+|h''_4|^2
+|h_7|^2
+\frac{1}{3}m^2_Z \cos 2 \beta \sin^2\theta_W, \nonumber\\
M^2_{33}&= M^2_{\tilde b_1}+\frac{v^2_1|y_1|^2}{2} +|h_4|^2 -\frac{1}{3}m^2_Z \cos 2 \beta \sin^2\theta_W, \nonumber\\
M^2_{44}&=  M^2_{\tilde Q}+\frac{v^2_2|y'_2|^2}{2} +|h_3|^2 +|h'_3|^2+|h''_3|^2
+|h_6|^2
 +m^2_Z \cos 2 \beta \left(\frac{1}{2}-\frac{1}{3}\sin^2\theta_W\right), \nonumber
\end{align}
\begin{align}
M^2_{55}&=M^2_{\tilde 2 L} +\frac{v^2_1|y_3|^2}{2} +|h'_3|^2 -m^2_Z \cos 2 \beta \left(\frac{1}{2}-\frac{1}{3}\sin^2\theta_W\right), \nonumber\\
M^2_{66}&= M^2_{\tilde b_2}+\frac{v^2_1|y_3|^2}{2}+|h'_4|^2  -\frac{1}{3}m^2_Z \cos 2 \beta \sin^2\theta_W,\nonumber\\
M^2_{77}&=M^2_{\tilde 3 L}+\frac{v^2_1|y_4|^2}{2}+|h''_3|^2-m^2_Z \cos 2 \beta \left(\frac{1}{2}-\frac{1}{3}\sin^2\theta_W\right),  \nonumber\\
M^2_{88}&= M^2_{\tilde b_3}+\frac{v^2_1|y_4|^2}{2}+|h''_4|^2   -\frac{1}{3}m^2_Z \cos 2 \beta \sin^2\theta_W\ . \nonumber\\
M^2_{99}&=M^2_{\tilde 4 L}+\frac{v^2_1|y_5|^2}{2}+|h_6|^2-m^2_Z \cos 2 \beta \left(\frac{1}{2}-\frac{1}{3}\sin^2\theta_W\right)\nonumber\\
M^2_{1010}&= M^2_{\tilde b_4}+\frac{v^2_1|y_5|^2}{2}+|h_7|^2   -\frac{1}{3}m^2_Z \cos 2 \beta \sin^2\theta_W\ . \nonumber\\
\end{align}

\begin{align}
M^2_{12}=M^{2*}_{21}=\frac{ v_2 y'_2h^*_3}{\sqrt{2}} +\frac{ v_1 h_4 y^*_1}{\sqrt{2}} ,
M^2_{13}=M^{2*}_{31}=\frac{y^*_1}{\sqrt{2}}(v_1 A^*_{b} -\mu v_2),
M^2_{14}=M^{2*}_{41}=0,\nonumber\\
 M^2_{15} =M^{2*}_{51}=h'_3 h^*_3,
 M^{2}_{16}= M^{2*}_{61}=0,  M^{2}_{17}= M^{2*}_{71}=h''_3 h^*_3,  M^{2}_{18}= M^{2*}_{81}=0,
M^{2}_{19}=M^{2*}_{91}=h^*_3 h_6,
\nonumber\\
M^{2}_{110}=M^{2*}_{101}=0,
M^2_{23}=M^{2*}_{32}=0,
M^2_{24}=M^{2*}_{42}=\frac{y'^*_2}{\sqrt{2}}(v_2 A^*_{B} -\mu v_1),  M^2_{25} = M^{2*}_{52}= \frac{ v_2 h'_3y'^*_2}{\sqrt{2}} +\frac{ v_1 y_3 {h'^*_4}}{\sqrt{2}} ,\nonumber\\
 M^2_{26}=M^{2*}_{62}=0,  M^2_{27} =M^{2*}_{72}=  \frac{ v_2 h''_3y'^*_2}{\sqrt{2}} +\frac{ v_1 y_4 h''^*_4}{\sqrt{2}},  M^2_{28} =M^{2*}_{82}=0, \nonumber\\
 M^2_{29} =M^{2*}_{92}=  \frac{ v_1 h^*_7y_5}{\sqrt{2}} +\frac{ v_2 y'^*_2 h_6}{\sqrt{2}},
M^{2}_{210}=M^{2*}_{102}=0,\nonumber\\
M^2_{34}=M^{2*}_{43}= \frac{ v_2 h_4 y'^*_2}{\sqrt{2}} +\frac{ v_1 y_1 h^*_3}{\sqrt{2}}, M^2_{35} =M^{2*}_{53} =0, M^2_{36} =M^{2*}_{63}=h_4 h'^*_4,\nonumber\\
 M^2_{37} =M^{2*}_{73} =0,  M^2_{38} =M^{2*}_{83} =h_4 h''^*_4,\nonumber\\
M^{2}_{39}=M^{2*}_{93}=0,
M^{2}_{310}=M^{2*}_{103}=h_4 h^*_7,\nonumber\\
M^2_{45}=M^{2*}_{54}=0, M^2_{46}=M^{2*}_{64}=\frac{ v_2 y'_2 h'^*_4}{\sqrt{2}} +\frac{ v_1 h'_3 y^*_3}{\sqrt{2}}, \nonumber\\
 M^2_{47} =M^{2*}_{74}=0,  M^2_{48} =M^{2*}_{84}=  \frac{ v_2 y'_2h''^*_4}{\sqrt{2}} +\frac{ v_1 h''_3 y^*_4}{\sqrt{2}},\nonumber\\
M^{2}_{49}=M^{2*}_{94}=0,
 M^2_{410} =M^{2*}_{104}=  \frac{ v_2 y'_2h^*_7}{\sqrt{2}} +\frac{ v_1 h_6 y^*_5}{\sqrt{2}},\nonumber\\
M^2_{56}=M^{2*}_{65}=\frac{y^*_3}{\sqrt{2}}(v_1 A^*_{s} -\mu v_2),
 M^2_{57} =M^{2*}_{75}=h''_3 h'^*_3,  \nonumber\\
 M^2_{58} =M^{2*}_{85}=0,
M^2_{59} =M^{2*}_{95}=h'^*_3 h_6,
 M^2_{510} =M^{2*}_{105}=0,
  M^2_{67} =M^{2*}_{76}=0,\nonumber\\
 M^2_{68} =M^{2*}_{86}=h'_4 h''^*_4,
 M^2_{69} =M^{2*}_{96}=0,
 M^2_{610} =M^{2*}_{106}=h'_4 h^*_7,
  M^2_{78}=M^{2*}_{87}=\frac{y^*_4}{\sqrt{2}}(v_1 A^*_{d} -\mu v_2)\ . \nonumber\\
 M^2_{79} =M^{2*}_{97}=h''^*_3 h_6,
 M^2_{710} =M^{2*}_{107}=0\nonumber\\
 M^2_{89} =M^{2*}_{98}=0,
 M^2_{810} =M^{2*}_{108}=h''_4 h^*_7,
  M^2_{910}=M^{2*}_{109}=\frac{y^*_5}{\sqrt{2}}(v_1 A^*_{b_4} -\mu v_2)\ . \nonumber
\label{14}
\end{align}
Next for $m^2_{ij}$ we have

\begin{align}
m^2_{11}&= M^2_{\tilde 1 L}+\frac{v^2_2|y'_1|^2}{2} +|h_3|^2 +m^2_Z \cos 2 \beta \left(\frac{1}{2}-\frac{2}{3}\sin^2\theta_W\right), \nonumber\\
m^2_{22}&=M^2_{\tilde T}+\frac{v^2_1|y_2|^2}{2}+|h_5|^2 +|h'_5|^2+|h''_5|^2
+|h_8|^2
 -\frac{2}{3}m^2_Z \cos 2 \beta \sin^2\theta_W, \nonumber\\
m^2_{33}&= M^2_{\tilde t_1}+\frac{v^2_2|y'_1|^2}{2} +|h_5|^2 +\frac{2}{3}m^2_Z \cos 2 \beta \sin^2\theta_W, \nonumber\\
m^2_{44}&=  M^2_{\tilde Q}+\frac{v^2_1|y_2|^2}{2} +|h_3|^2 +|h'_3|^2+|h''_3|^2
+|h_6|^2
 -m^2_Z \cos 2 \beta \left(\frac{1}{2}-\frac{2}{3}\sin^2\theta_W\right), \nonumber
\end{align}
\begin{align}
m^2_{55}&=M^2_{\tilde 2 L} +\frac{v^2_2|y'_3|^2}{2} +|h'_3|^2 +m^2_Z \cos 2 \beta \left(\frac{1}{2}-\frac{2}{3}\sin^2\theta_W\right), \nonumber\\
m^2_{66}&= M^2_{\tilde t_2}+\frac{v^2_2|y'_3|^2}{2}+|h'_5|^2  +\frac{2}{3}m^2_Z \cos 2 \beta \sin^2\theta_W,\nonumber\\
m^2_{77}&=M^2_{\tilde 3 L}+\frac{v^2_2|y'_4|^2}{2}+|h''_3|^2+m^2_Z \cos 2 \beta \left(\frac{1}{2}-\frac{2}{3}\sin^2\theta_W\right),  \nonumber\\
m^2_{88}&= M^2_{\tilde t_3}+\frac{v^2_2|y'_4|^2}{2}+|h''_5|^2   +\frac{2}{3}m^2_Z \cos 2 \beta \sin^2\theta_W,\nonumber\\
m^2_{99}&=M^2_{\tilde 4 L}+\frac{v^2_2|y'_5|^2}{2}+|h_6|^2+m^2_Z \cos 2 \beta \left(\frac{1}{2}-\frac{2}{3}\sin^2\theta_W\right),  \nonumber\\
m^2_{1010}&= M^2_{\tilde t_4}+\frac{v^2_2|y'_5|^2}{2}+|h_8|^2   +\frac{2}{3}m^2_Z \cos 2 \beta \sin^2\theta_W.\
\nonumber
\end{align}

\begin{align}
m^2_{12}&=m^{2*}_{21}=-\frac{ v_1 y_2h^*_3}{\sqrt{2}} +\frac{ v_2 h_5 y'^*_1}{\sqrt{2}} ,
m^2_{13}=m^{2*}_{31}=\frac{y'^*_1}{\sqrt{2}}(v_2 A^*_{t} -\mu v_1),
m^2_{14}=m^{2*}_{41}=0,\nonumber\\
 m^2_{15} &=m^{2*}_{51}=h'_3 h^*_3,
 m^{2}_{16}= m^{2*}_{61}=0,  m^{2*}_{17}= m^{2*}_{71}=h''_3 h^*_3,  m^{2*}_{18}= m^{2*}_{81}=0,\nonumber\\
m^2_{23}&=m^{2*}_{32}=0,
m^2_{24}=m^{2*}_{42}=\frac{y^*_2}{\sqrt{2}}(v_1 A^*_{T} -\mu v_2),  m^2_{25} = m^{2*}_{52}= -\frac{ v_1 h'_3y^*_2}{\sqrt{2}} +\frac{ v_2 y'_3 h'^*_5}{\sqrt{2}} ,\nonumber\\
 m^2_{26} &=m^{2*}_{62}=0,  m^2_{27} =m^{2*}_{72}=  -\frac{ v_1 h''_3y^*_2}{\sqrt{2}} +\frac{ v_2 y'_4 h''^*_5}{\sqrt{2}},  m^2_{28} =m^{2*}_{82}=0, \nonumber\\
m^2_{34}&=m^{2*}_{43}= \frac{ v_1 h_5 y^*_2}{\sqrt{2}} -\frac{ v_2 y'_1 h^*_3}{\sqrt{2}}, m^2_{35} =m^{2*}_{53} =0, m^2_{36} =m^{2*}_{63}=h_5 h'^*_5,\nonumber\\
 m^2_{37} &=m^{2*}_{73} =0,  m^2_{38} =m^{2*}_{83} =h_5 h''^*_5,\nonumber\\
m^2_{45}&=m^{2*}_{54}=0, m^2_{46}=m^{2*}_{64}=-\frac{y'^*_3 v_2 h'_3}{\sqrt{2}}+\frac{v_1 y_2 h'^*_5}{\sqrt{2}},
\nonumber\\
m^2_{47}&=m^{2*}_{74}=0,
m^2_{48}=m^{2*}_{84}=\frac{v_1 y_2 h''^*_5}{\sqrt{2}}-\frac{v_2 y'^*_4 h''_3}{\sqrt{2}},\nonumber\\
 m^2_{56}&=m^{2*}_{65}=\frac{y'^*_3}{\sqrt{2}}(v_2 A^*_{c}-\mu v_1), \nonumber\\
m^2_{57}&=m^{2*}_{75}= h''_3 h'^*_3, m^2_{58}=m^{2*}_{85}=0, \nonumber\\
m^2_{67}&=m^{2*}_{76}=0, m^2_{68}=m^{2*}_{86}= h'_5 h''^*_5, \nonumber\\
m^2_{78}&=m^{2*}_{87}=\frac{y'^*_4}{\sqrt{2}}(v_2 A^*_{u}-\mu v_1),\nonumber\\
m^2_{19}&=m^{2*}_{91}=h_6 h^*_3, m^2_{110}=m^{2*}_{101}=0, \nonumber\\
m^2_{29}&=m^{2*}_{92}=-\frac{y^*_2 v_1 h_6}{\sqrt{2}}+\frac{v_2 y^*_5 h_8}{\sqrt{2}},\nonumber\\
m^2_{210}&=m^{2*}_{102}=0, m^2_{39}=m^{2*}_{93}=0,\nonumber\\
m^2_{310}&=m^{2*}_{103}=h_5 h^*_8,\nonumber\\
m^2_{49}&=m^{2*}_{94}=0, m^2_{410}=m^{2*}_{104}=
-\frac{y'^*_5 v_2 h_6}{\sqrt{2}}+\frac{v_1 y_2 h^*_8}{\sqrt{2}},\nonumber\\
m^2_{59}&=m^{2*}_{95}=h_6 h'^*_3, m^2_{510}=m^{2*}_{105}=0\nonumber\\
m^2_{69}&=m^{2*}_{96}=0, m^2_{610}=m^{2*}_{106}= h'_5 h^*_8 \nonumber\\
m^2_{79}&=m^{2*}_{97}=h_6 h''^*_3, m^2_{710}=m^{2*}_{107}=0, \nonumber\\
m^2_{89}&=m^{2*}_{98}=0, m^2_{810}=m^{2*}_{108}=h''_5 h^*_8, \nonumber\\
 m^2_{910}&=m^{2*}_{109}=\frac{y'^*_5}{\sqrt{2}}(v_2 A^*_{t_4}-\mu v_1)
\end{align}

\section{{Appendix B:} CP even-CP odd Higgs mixing with vector like quarks \label{sec8}}
For completeness we give here a brief  discussion of  the mixings of CP even Higgs and 
CP odd Higgs which arise as a consequence of CP phases in the soft SUSY parameters of the
theory. While there is no CP violation in the Higgs sector at the tree level, the CP phases
from the soft breaking sector induce CP violation at the loop level. Thus at the tree level the 
scalar potential in the Higgs sector is given by 
\begin{align}
V_0&=m_1^2 |H_1|^2+m_2^2|H_2|^2 +(m_3^2 H_1.H_2 + H.C.)
+\frac{(g_2^2+g_1^2)}{8}|H_1|^4\non
&+
\frac{(g_2^2+g_1^2)}{8}|H_2|^4
-\frac{g_2^2}{2}|H_1.H_2|^2
+\frac{(g_2^2-g_1^2)}{4}|H_1|^2|H_2|^2\,.
\end{align}
This potential has no CP violation. However, there are important loop corrections to the potential.
At the one loop level they are  given by 
\beq
\Delta V=\frac{1}{64\pi^2}
 Str(M^4(H_1,H_2)(log\frac{M^2(H_1,H_2)}{Q^2}-\frac{3}{2}))\,,
\eeq
where the super trace sums over bosons and fermions circulating in the loop.
In our case the largest contributions  to the scalar potential arise from the third generation 
quarks and squarks, and from the vector like quarks and their super partners in the loop.

The loop corrections introduce  a CP phase in the Higgs sector and one may parametrize the
 Higgs fields so that 

\beqn
(H_1)= \left(\begin{matrix} H_1^0\cr
 H_1^-  \end{matrix}  \right)
 =
\left(\begin{matrix} \frac{1}{\sqrt 2}(v_1+\phi_1+i\psi_1)\cr
             H_1^-\end{matrix}\right)\,,
    \eeqn
     \beqn
 (H_2)= \left(\begin{matrix}H_2^+\cr
             H_2^0\end{matrix}\right)
=e^{i\theta_H} \left(\begin{matrix}H_2^+ \cr
            \frac{1}{\sqrt 2} (v_2+\phi_2+i\psi_2)\end{matrix}\right)\,.
\eeqn
The minimization of the scalar potential including the loop correction induces mixing of CP even 
and CP odd Higgs fields. Thus at the tree level one may write 
 the components of the neutral Higgs fields 
  as  $\Phi_a$ (a=1-4) where $\Phi_a=(\phi_1, \phi_2, \psi_1,\psi_2)$ and where
$\phi_1, \phi_2$ are CP even and $\psi_1,\psi_2$ are CP odd. 
After minimization of the full potential including the loop corrections we find a mixing of CP even
 and CP odd states  and the mass matrix in the neutral Higgs sector then takes the following form 

\beq
M_{ab}^2= M_{ab}^{2(0)}+ \Delta M_{ab}^2\,.
\label{mass-loop}
\eeq
Here  $M_{ab}^{2(0)}$ are the contributions at the tree level and
 $\Delta M_{ab}^2$ are the contributions at the loop level. The 
 loop correction to the mass squared matrix takes the form 

\beq
\Delta M_{ab}^2=
\frac{1}{32\pi^2}
Str(\frac{\partial M^2}{\partial \Phi_a}\frac{\partial M^2}{\partial\Phi_b}
log\frac{M^2}{Q^2}+M^2 \frac{\partial^2 M^2}{\partial \Phi_a\partial \Phi_b}
log\frac{M^2}{eQ^2})_0\,,
\eeq
 where e=2.718.  Eq.(38) gives a $4\times 4$ mass square matrix in the basis
 $(\phi_1, \phi_2, \psi_1, \psi_2)$.  The $4\times 4$ mass squared matrix can be reduced to a 
 $3\times 3$ mass squared matrix  by use of the 
following linear combinations of $\psi_1$ and $\psi_2$.
\begin{align}
\psi'_{1}&=\sin\beta \psi_1 + \cos\beta \psi_2\,,\nonumber\\
\psi'_{2}&=-\cos\beta\psi_1 +\sin\beta \psi_2\,,
\end{align}
where $\tan\beta = <H_2>/<H_1>$. 
In the new basis $\psi'_{2}$ decouples and can be identified  as a Goldstone field with a zero mass eigenvalue.
The remaining  Higgs mass squared  matrix  now  involves only three fields and in the basis $(\phi_1, \phi_2, \psi_1')$ 
 is given by
\beq
M^2_{Higgs}=
\left(\begin{matrix}M_Z^2c_{\beta}^2+M_A^2s_{\beta}^2+\Delta_{11} &
-(M_Z^2+M_A^2)s_{\beta}c_{\beta}+\Delta_{12} &\Delta_{13}\cr
-(M_Z^2+M_A^2)s_{\beta}c_{\beta}+\Delta_{12} &
M_Z^2s_{\beta}^2+M_A^2c_{\beta}^2+\Delta_{22} & \Delta_{23} \cr
\Delta_{13}  &\Delta_{23} &(M_A^2+\Delta_{33}) \end{matrix}\right)\,.
\eeq
Here $s_{\beta}= \sin\beta$, $c_{\beta} = \cos\beta$, $M_Z$ is the $Z$-boson mass, and $M_A$ is the 
mass of the CP odd Higgs before mixing. 
As mentioned in section \ref{sec2} a  detailed analysis of $\Delta_{ij}$ is given in~\cite{Ibrahim:2016rcb} 
including a vectorlike quark generation 
and we  utilize the results of ~\cite{Ibrahim:2016rcb}  
 in the computation of the Higgs mass eigenstates and mixings in the analysis given in this work.

\end{document}